\documentclass[twocolumn]{aastex631}
\usepackage{placeins}

\usepackage{lmodern}
\usepackage{latexsym}
\usepackage{amsmath}
\usepackage{amssymb}
\usepackage{amsbsy}
\usepackage{amsthm}
\usepackage{amsfonts}
\usepackage{mathrsfs}
\usepackage{bm}
\usepackage{sansmath}
\usepackage{relsize}
\usepackage{caption2}
\usepackage{graphicx}
\usepackage[utf8]{inputenc} 
\usepackage[T1]{fontenc}
\usepackage{epstopdf}
\usepackage{esdiff}
\usepackage[T1]{fontenc}
\usepackage{ae,aecompl}
\usepackage{mathtools}
\usepackage{array}
\usepackage{booktabs}
\usepackage{longtable}
\usepackage{appendix}
\usepackage{hyperref}
\usepackage{acronym}
\usepackage{natbib}
\usepackage{tikz}

\shorttitle{}
\shortauthors{Ranjbar, Olivares, Abbassi}
\graphicspath{{./}{figures/}}
\begin{document}

\title{Bondi-like Accretion Flow Dynamics: The Role of Gravitational Potential}

\email{r.ranjbar@ipm.ir, h.sanchez@ua.pt, sabbassi@uwo.ca}

\author[0000-0003-1488-4890]{Razieh Ranjbar}
\affiliation{School of Astronomy, Institute for Research in Fundamental Sciences (IPM), P.O. Box 19395-5531, Tehran, Iran}

\author[0000-0001-6833-7580]{Héctor R. Olivares-Sánchez}
\affiliation{ Departamento de Matem\'{a}tica da Universidade de Aveiro and Centre for Research and Development in Mathematics and Applications (CIDMA), Campus de Santiago, 3810-193 Aveiro, Portugal}

\author[0000-0003-0428-2140]{Shahram Abbassi}
\affiliation{Department of Physics \& Astronomy, Western University, London, Canada}

\newacro{AGN}{active galactic nuclei}
\newacro{XRB}{X-ray binaries}
\newacro{BH}{black hole}
\newacro{SMBH}{super massive black hole}
\newacro{GR}{general relativistic}
\newacro{DM}{dark matter}




\begin{abstract}

The formation of massive \acp{BH} and their coevolution with host galaxies are pivotal areas of modern astrophysics. Spherical accretion onto a central point mass serves as a foundational framework in cosmological simulations, semi-analytical models, and observational studies. In this paper, we extend the classical spherical accretion model by incorporating the gravitational potential of host galaxies, including contributions from stellar components and \ac{DM} halos. Numerical solutions spanning scales from parsecs down to $\sim 10 \ r_{\rm s}$ reveal that the flow structure is highly sensitive to the mass and size of the \ac{DM} halo. Adding a small amount of angular momentum to the accreting gas demonstrates that such flows resemble spherical Bondi accretion, with mass accretion rates converging toward the Bondi rate. We find that the low angular momentum flow resembles the spherical Bondi flow, and its mass accretion rate approaches the Bondi accretion rate. Remarkably, due to the presence of \ac{DM}, the mass accretion rate increases by more than  $\sim \% 100$ compared to analogous hydrodynamic solutions without dark matter. These findings underscore the critical role of stellar and \ac{DM} gravitational potentials in shaping the dynamics and accretion rates of quasi-spherical flows, providing new insights into astrophysical accretion processes.

\end{abstract}

\keywords{galaxies: evolution --- black hole physics --- accretion, accretion discs --- ISM: jets and outflows --- hydrodynamics}

\section{Introduction} \label{sec:intro}

Almost every galaxy hosts a \ac{SMBH}. Black hole accretion systems, such as \acp{XRB} or \acp{AGN}, can release substantial amounts of energy into their surrounding interstellar environments. The mass of a black hole is closely correlated with various properties of its host galaxy, indicating a co-evolution between the central \ac{BH} and its host galaxy \cite{2002ApJ...574..740T, 2013ARAA..51..511K}. However, the mechanisms by which black holes are coupled to larger-scale structures, including their host \ac{DM} halos, remain poorly understood. Several processes have been proposed that drive gas into the galaxy's center, eventually triggering \ac{AGN} activity, and these processes may strongly depend on either the cosmic environment or the properties of the \ac{DM} halos hosting the galaxies \cite{2008ApJS..175..356H}. Understanding these mechanisms is crucial for explaining galaxy cluster formation, particularly for low-luminosity \acp{AGN}, whose accretion dynamics often deviate significantly from standard accretion scenarios.

In recent decades, numerous analytical models have been developed to elucidate the fundamental physical processes governing the accretion flow. The simplest model for accretion is spherical accretion, which assumes a spherically symmetric inflow \cite{1952MNRAS.112..195B}. While this model serves as a valuable theoretical framework, it represents an idealized scenario. In reality, angular momentum, even if minimal, plays a key role in determining accretion dynamics. Deviations from idealized spherical flows, particularly in the presence of small angular momentum or environmental complexities, remain an important area of research that has been taken into consideration of several recent studies (e.g., \citep{2023ApJ...945...76H, 2022MNRAS.516.3984R, 2011MNRAS.415.3721N};).

The classical Bondi problem has been instrumental in analyzing accretion onto primordial black holes in the early universe, before the formation of galaxies and stars \cite{2007ApJ...662...53R}. The central \ac{AGN} is typically located in the densest region of a cluster's hot atmosphere. Consequently, classical Bondi accretion of hot gas is often regarded as one of the primary mechanisms fueling central black holes. \cite{2011MNRAS.415.3721N} introduced an accretion flow known as the slowly rotating accretion flow. This model serves as an intermediate case between Bondi accretion and the advection-dominated accretion flow (ADAF), making it particularly relevant to the central regions of elliptical galaxies. Moreover, fully general relativistic simulations and analytical solutions of spherical accretion (e.g., \citep{2020ApJ...893...81T, 2023AA...678A.141O, 2025AA...RH, 2024MPLA...3950076D, 2024JCAP...09..006D}) reveal key features, offering deeper insights into their applicability and limitations under conditions of strong gravity. Incorporating additional physics, such as time-dependent processes or magnetohydrodynamic (MHD) effects, could enhance the predictive power and relevance of these models to \ac{AGN} observations.

A spherical or quasi-spherical flow has no angular momentum, so the loss of angular momentum is not essential for accretion. Bondi flows with purely spherical surfaces or flows with angular momentum below the minimum value required for stable circular orbits around black holes are examples of such solutions. Recent investigations of Bondi accretion and quasi-spherical accretion have given considerable attention to the influence of the gravitational potential exerted by the stars within the galaxy \cite{2017ApJ...848...29C, 2018ApJ...868...91C, 2019MNRAS.489.3870S, 2023ApJ...954..117R}. The presence of central black holes, gas, and stars in the universe is well established through direct observations. However, our knowledge of the nature of \ac{DM} remains very limited. While its presence is strongly suggested by gravitational effects and widely used in cosmological models, it has yet to be directly detected. Ongoing research may one day confirm its existence or uncover an alternative explanation for the effects currently attributed to this component. However, as our solution involves large-scale distances, the potential impact of dark matter cannot be overlooked. In this study, we examine
how a spherically symmetric \ac{DM} halo, modeled
using a Hernquist profile, modifies the dynamics and
structure of Bondi-like accretion flows. The analysis is
conducted within a Newtonian framework using semi-analytical methods.

 The classical Bondi solution is used in the interpretation of observational results, in numerical investigations, and cosmological simulations involving galaxies and accretion on their central \acp{BH}. In many such studies, when the instrumental resolution is limited, or the numerical resolution is inadequate, an estimate of the mass accretion rate is derived using the classical Bondi solution, taking values of temperature and density measured at some finite distance from the \ac{BH}. This procedure clearly produces an estimate that can depart from the true value, even when assuming that accretion fulfills the assumptions of the Bondi model (stationarity, spherical symmetry, etc.). \citep{2022MNRAS.516.3984R, 2023ApJ...954..117R} developed the analytical setup of the problem for generic accretion from classical Bondi accretion by including additional physical effects such as galactic potential and slow rotation and outflow; they also numerically investigated the deviation of the mass accretion rate from the classical Bondi rate in the presence of rotation. Here we reconsider the problem specifically for the Bondi-like case in the presence of a \ac{DM} halo, an aspect that was not discussed in detail in \cite{2022MNRAS.516.3984R, 2023ApJ...954..117R}, and we also extend the investigation to the case of the Hernquist potential.

 In addition to these considerations, the gravitational potential of the host galaxy, which includes both stellar and dark matter
components of galaxies, must also be taken into account when studying accretion flows on scales beyond the immediate vicinity of black holes. Estimating the influence of the host galaxy's gravitational potential requires complex calculations, a factor that is frequently overlooked in previous research.

Many galaxies are known to be enveloped by \ac{DM}, and it is reasonable to expect that \acp{BH} and their accretion flows are similarly surrounded by \ac{DM} halos. The distribution of \ac{DM} can vary depending on the mass and type of the galaxy. Among the various profiles used to describe dark matter, the NFW (Navarro-Frenk-White) \cite{1996ApJ...462..563N}, Hernquist \cite{1990ApJ...356..359H}, Jaffe \cite{1983MNRAS.202..995J}, and King \cite{1962AJ.....67..471K} profiles are widely recognized. For the present study, we specifically consider the Hernquist profile, as it provides a simple and spherical model, offering a straightforward representation of dark matter in elliptical galaxies. The gravitational response of black holes within realistic astrophysical environments has previously been explored using this model \cite{2022PhRvD.105f1501C}.

Observations indicate that the distribution of both the stellar (baryonic matter) and \ac{DM} mass profile within galaxies becomes significant at large radial distances (e.g., \citep{2002ApJ...575...87T, 2004ApJ...611..739T, 2007ApJ...667..176G}). In practice, physical processes occurring on sub-pc and pc scales can be quite complex. Observations show an \ac{SMBH} of $ \sim 10^8 - 10^9 M_{\rm \odot}$ is embedded in a stellar bulge of approximately a few hundred parsecs in size, e.g., \cite{1999AA...348..768P}. Close to the \ac{BH}, its gravitational potential drastically influences the movement of stars and the interstellar medium. However, at larger distances, the gravitational impact of the \ac{BH} on the overall dynamics of the stellar and interstellar media decreases significantly. In such scenarios, alongside the \ac{BH}'s potential, the gravitational potential of the galaxy stars becomes consequential on the parsec scales and must be taken into account. At larger scales, such as the galaxy scale, it is essential to incorporate the gravitational potential of host galaxies \cite{2018MNRAS.478.2887Y}, \cite{2016ApJ...818...83B}, \cite{2021MNRAS.505.4129S}. Previous studies have explored Bondi accretion at the galaxy scale, considering the influence of the host galaxy's gravitational potential (\citep{2011MNRAS.418..591B, 2016MNRAS.460.1188K, 2017ApJ...848...29C}). In addition to the stellar gravitational potential, the presence of dark matter can significantly affect the structure of the accretion flow. \cite{2022MNRAS.512.2474M} studied Bondi accretion in two-component galaxy models, adopting a Jaffe profile for the stellar distribution embedded within a dark matter halo following an NFW profile at large radii. In a similar spirit, in this work we investigate low-angular-momentum accretion flows under the combined influence of a Hernquist-profile dark matter halo and the stellar gravitational potential, which distinguishes our study from previous ones.

In this study, our aim is to systematically investigate how adding the gravitational potentials of a \ac{DM} halo and stellar bulge to the equation of motion of Bondi accretion impacts its structure and mass accretion rate. Additionally, we explore the influence of the \ac{DM} potential on a slowly rotating accretion flow. Ultimately, we seek to understand the combined effect of low angular momentum and \ac{DM} potential on spherical accretions. By employing more realistic gravitational potentials, we aim to explore their effects on the characteristics of different accretion flow solutions. In our numerical solutions, the outer boundary is set precisely at the Bondi radius.

The paper is organized as follows. In section \ref{sec.2}, we consider the generalized case of Bondi-like accretion in the presence of a galactic gravitational potential hosting the central \ac{BH}. Section \ref{sec.3} presents the equations for accretion flow with low angular momentum. In Section \ref{sec.4}, we present our numerical results and analysis. Finally, in Section \ref{sec.5}, we discuss the implications and key findings of our results, with additional technical details provided in the appendix.

\begin{figure*}
\centering
\makebox[\linewidth]{%
\begin{tabular}{cc}
    \includegraphics[width=0.4\linewidth]{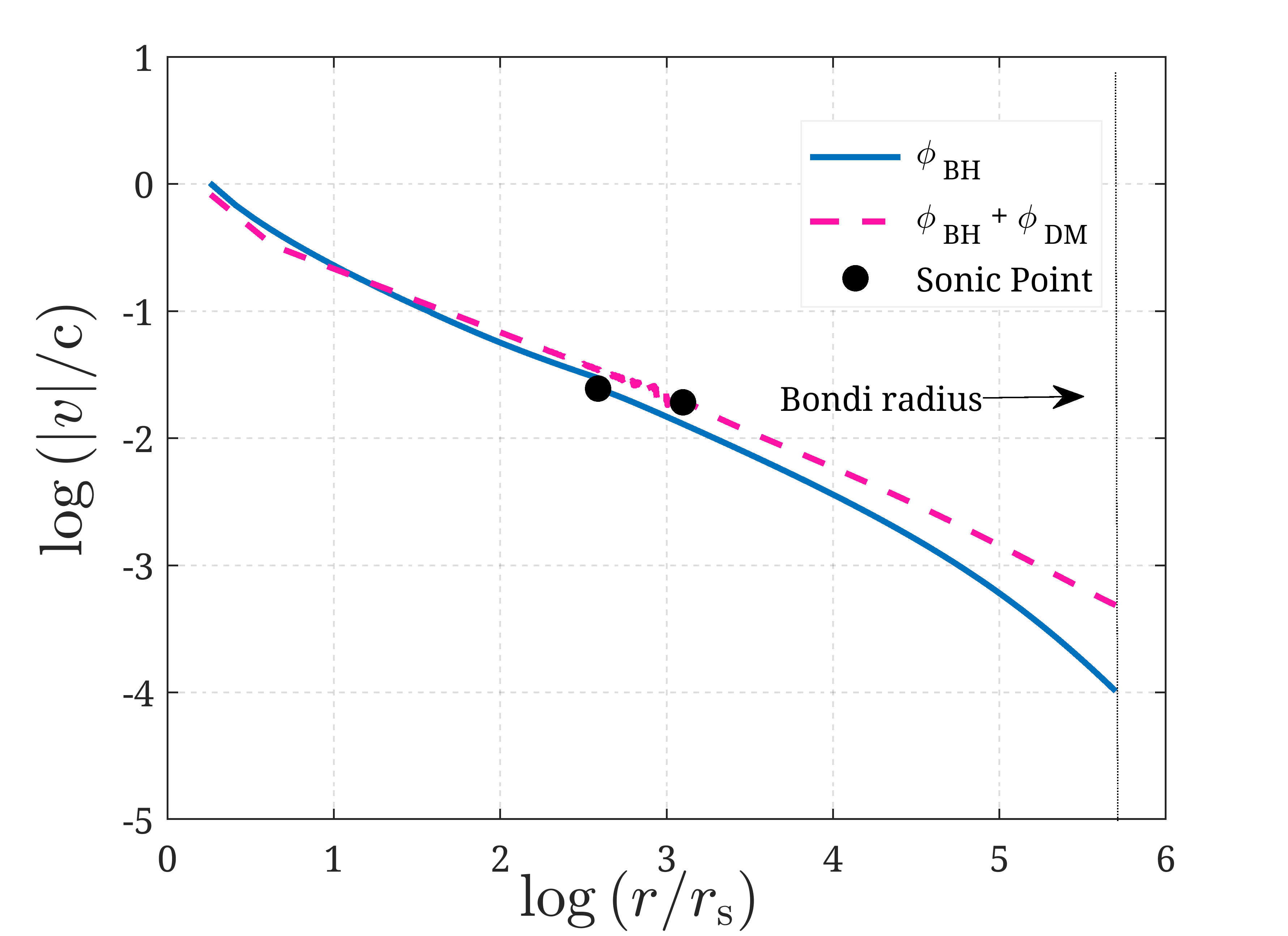}&
    \includegraphics[width=0.4\linewidth]{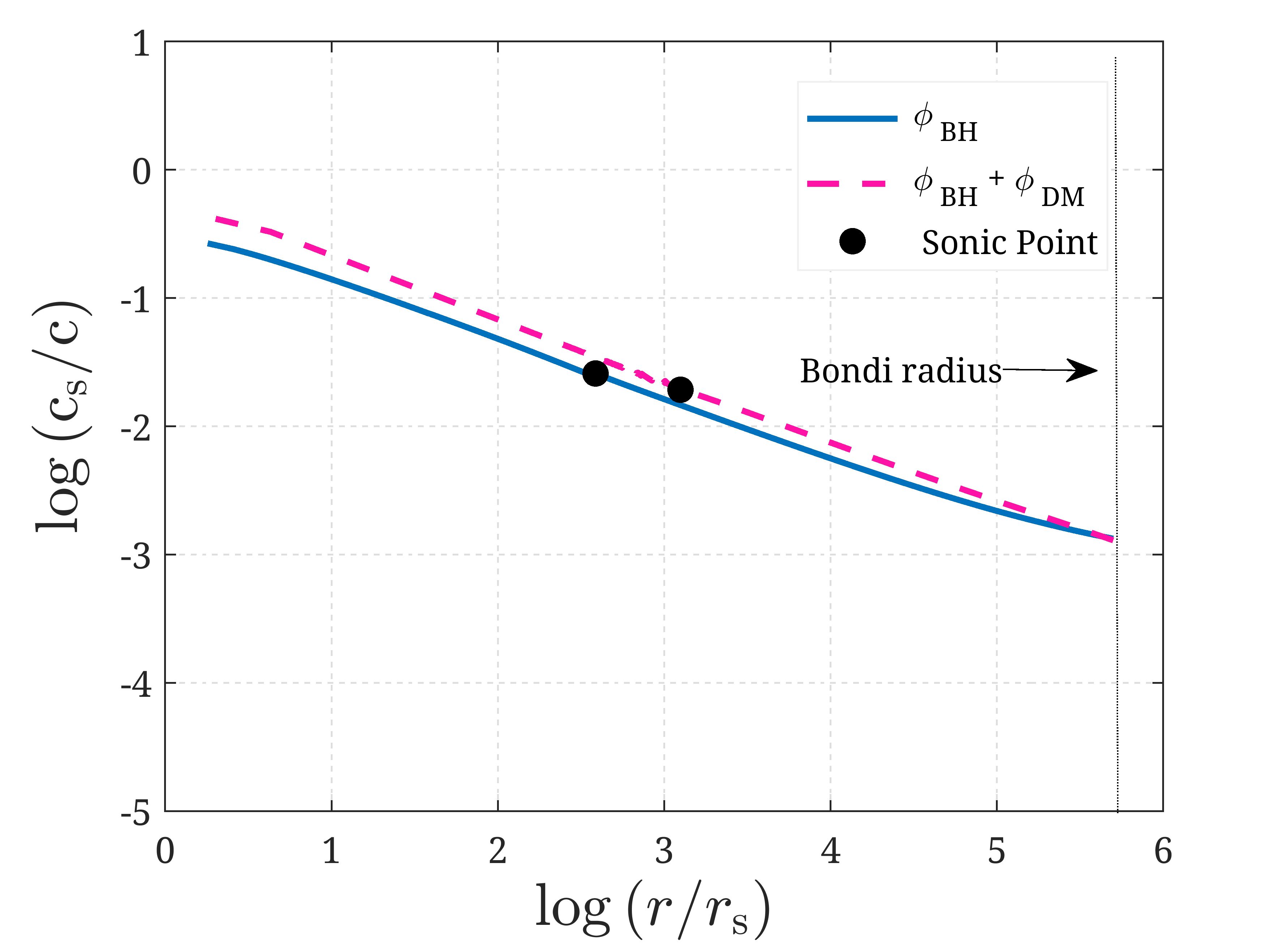}\\
    \includegraphics[width=0.4\linewidth]{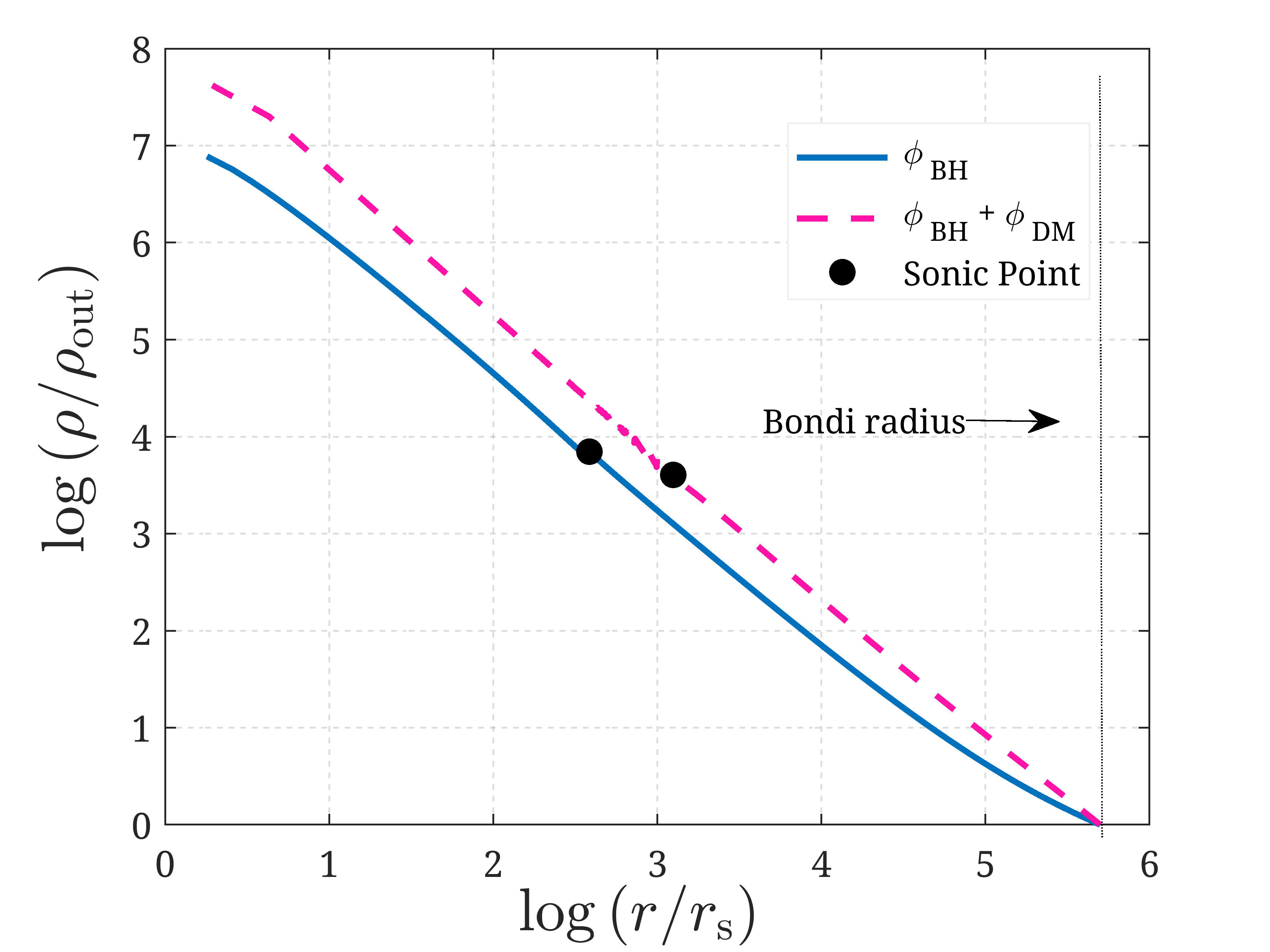}&
    \includegraphics[width=0.4\linewidth]{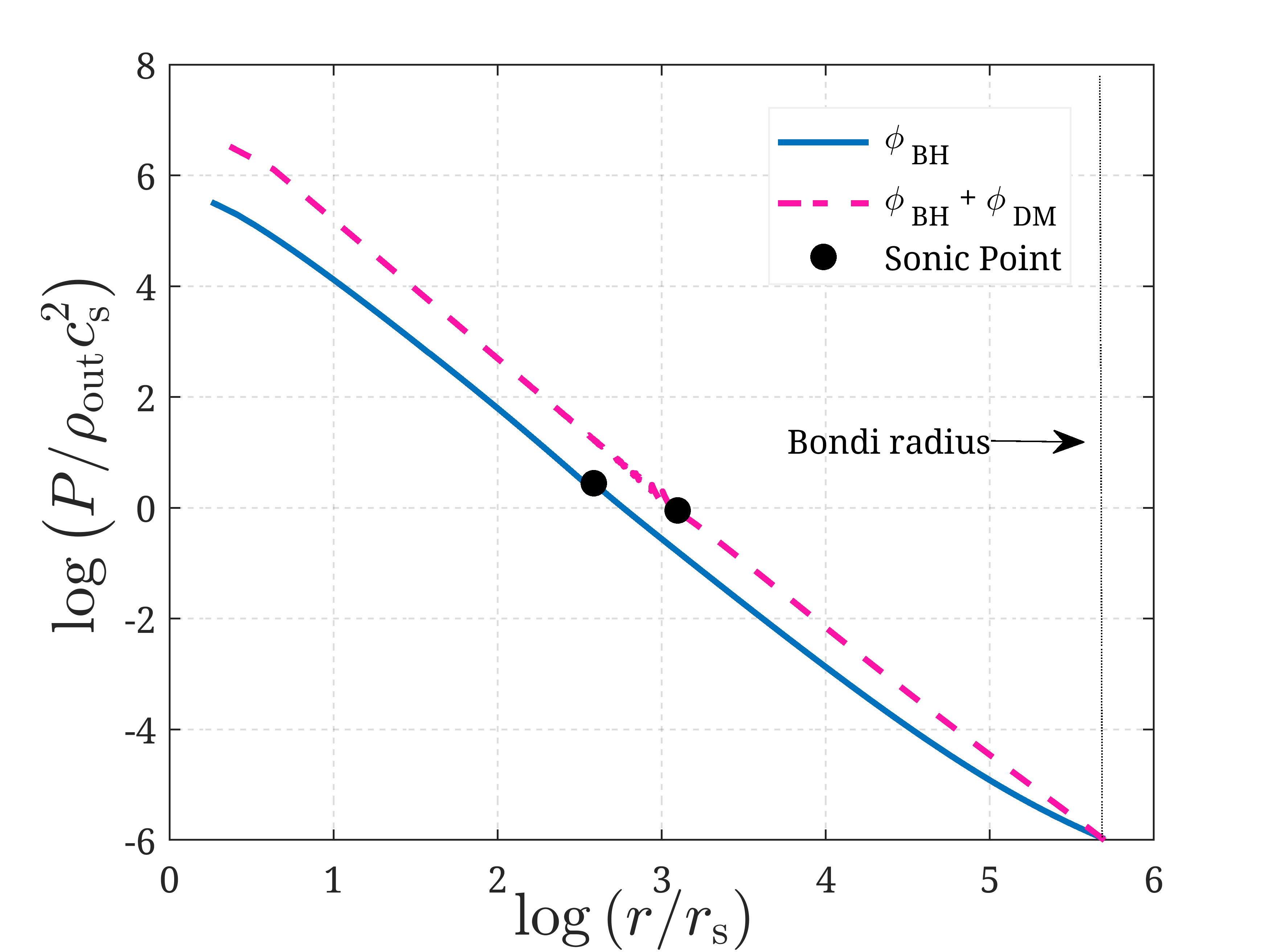}
  \end{tabular}%
  }
  \centering  
    \caption{Global solutions for Bondi accretion, $\gamma = 5/3$,\ $M_{\rm halo} = M_{\rm BH}$, $a_0 = 100$. Blue line related without dark matter potential,($\phi_{\rm DM} = 0,\ r_c = 386.74$). Dashed pink line ($\phi_{\rm DM} \neq 0,\ r_{\rm c} = 1253.6$). The galaxy's stellar potential is $\phi_{\rm s} \neq 0$. A solid dot indicates the position of the critical point. The vertical dotted lines correspond to the location of the Bondi radius $r_{\rm B}$.}
    \label{1}
\end{figure*}

\section{Bondi-like Flow}\label{sec.2}

The mechanisms by which \acp{SMBH} in galactic nuclei are fed remain uncertain. For spherically symmetric accretion of hot gas \cite{1952MNRAS.112..195B}, accretion typically begins from a characteristic radius, known as the Bondi radius, defined as:

\begin{equation}
r_B = \frac{2GM_{\rm BH}}{c_{\rm s,\infty}^2},
\end{equation}
where $G$ is the gravitational constant, $M_{\rm BH}$ is the black hole mass, and $c_{\rm s,\infty}$ is the sound speed of the ambient gas. At this radius, the gravitational energy due to the \ac{SMBH} becomes comparable to the thermal energy of the gas. The Bondi radius represents the scale
at which the \ac{SMBH}’s gravitational influence transitions from negligible to dominant, offering a natural framework for modeling spherical accretion flows.

In this section, we analyze the dynamics of accreting matter in the combined gravitational potential of a host galaxy and a central black hole, assuming a steady-state, spherically symmetric flow (i.e., no angular momentum). We consider the simplest case: spherically symmetric steady accretion under the gravitational field of a \ac{BH} plus host galaxy (including both stellar and \ac{DM} components). The inclusion of both stellar and \ac{DM} gravitational potentials allows for a more comprehensive understanding of accretion dynamics at scales beyond the immediate vicinity of the \ac{SMBH}, bridging the gap between galactic and sub-galactic scales. Spherical accretion onto a gravitating body was first studied by \citet{1952MNRAS.112..195B} and is often called Bondi accretion, which is now believed to be quite significant. While the classical Bondi model is invaluable for its simplicity and mathematical elegance, it has limitations when applied to realistic astrophysical scenarios. These include neglecting the influence of large-scale gravitational potentials, angular momentum, and potential radiative feedback effects. By extending the classical Bondi framework to incorporate these complexities, we can substantially improve its relevance to modern observations and numerical simulations.
Our analysis incorporates the combined gravitational potential of the black hole, stellar bulge, and \ac{DM} halo. This multi-component approach allows us to study how each component contributes to the accretion dynamics, with a focus on large-scale astrophysical processes. The solutions derived here provide a baseline for understanding more complex and realistic accretion scenarios, including rotating and radiatively inefficient accretion flows.

\subsection{Basic Equations}

We consider a spherically symmetric flow around a black hole embedded in a \ac{DM} halo. The flow is assumed to be steady and one-dimensional in the radial ($r$) direction. We neglect the magnetic field and radiation effects, and the flow is further assumed to be inviscid and adiabatic. Under the Newtonian approximation, the continuity equation and the equation of motion governing this infalling matter are given by:

\begin{equation}
\frac{1}{4\pi r^2}\frac{d}{dr}(\rho v r^2) = 0,
\end{equation}

\begin{equation}
v \frac{dv}{dr} = - \frac{d\Phi}{dr} - \frac{1}{\rho} \frac{dp}{dr},
\end{equation}

where $\Phi$ is the gravitational potential (for more details, see \ref{GP}), $v$ is the flow velocity (negative for accretion), $\rho$ is the density, and $p$ is the pressure. These equations form the foundation for describing spherical accretion in the combined gravitational field of the black hole and the host galaxy, allowing us to explore the impact of stellar and \ac{DM} potentials on the flow dynamics.

\begin{figure*}
\centering
70    \includegraphics[width=0.7\linewidth]{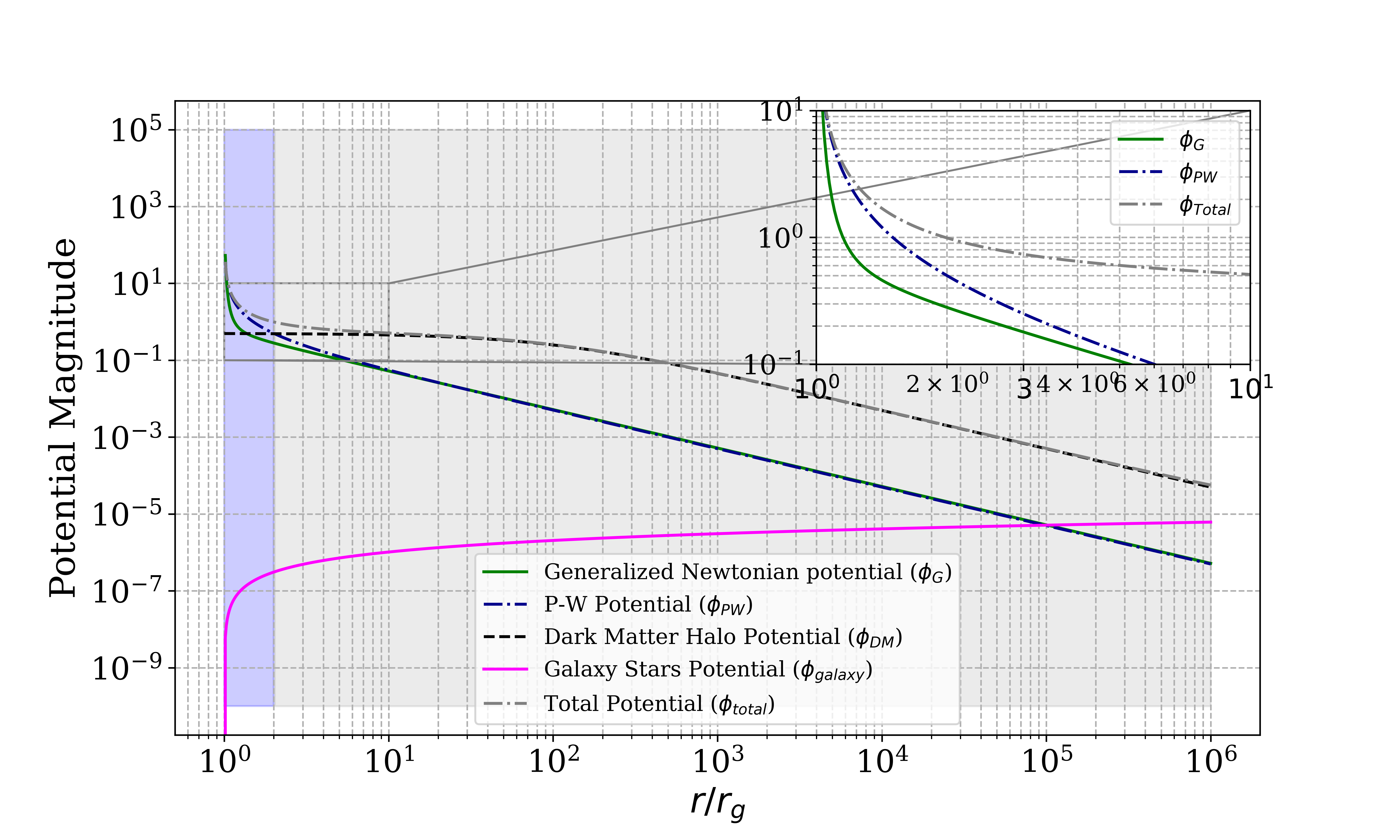} 
  \centering  
    \caption{Dominance Regions of Gravitational Potentials. The magnitudes of the potentials for the black hole (Generalized Newtonian $\phi_{\rm G}$, Paczynski \& Wiita $\phi_{\rm PW}$), dark matter halo, and galaxy stars are plotted at different length scales. The dot-dashed grey line is their sum. In our normalized units ($r_s = c = 1$), gravitational potentials are dimensionless. The small panel provides a zoomed-in view of the central region. Shaded regions are added to indicate where each potential dominates over the others. Blue shade for the black hole potential dominance. Grey shade for the dark matter potential dominance.}
    \label{P}
\end{figure*}

Integration of the continuity equation gives:

\begin{equation}
-4\pi \rho v r^2 = \dot{M},
\end{equation}
where $\dot{M}$ is the mass accretion rate, which is constant in this case. This result implies a steady inflow of mass, with the accretion rate governed by the density and velocity profiles of the gas. To solve this equation, we follow the approach outlined in previous studies.
It is important to clarify that in Bondi accretion, the entropy of the gas is not generally conserved for an arbitrary polytropic index $\gamma$. Entropy conservation holds only in the special case of smooth, adiabatic flows with no shocks or radiative losses. In this part of our work, we restrict to this case and, following \cite{2011MNRAS.415.3721N}, adopt the isothermal definition of the sound speed $c_{\rm s}^2 = p/\rho$, so that

\begin{equation}
S = \frac{c_s^2}{\rho^{\gamma - 1}} = \text{constant},
\end{equation}
where $S$ is a conserved quantity related to the entropy, that can be derived from the energy equation, and $\gamma$ is the ratio of specific heats of the gas, set to $\gamma = 5/3$ \citep[c.f.][]{2011MNRAS.415.3721N}. Consequently, the Bernoulli equation is given by:

\begin{equation}
\frac{1}{2} v^2 + \Phi + \frac{\gamma S}{(\gamma - 1)r^{2(\gamma-1)} v^{\gamma-1}} \left(\frac{\dot{M}}{4\pi}\right)^{\gamma -1} = B,
\end{equation}
where $B$ is the Bernoulli constant. The Bernoulli equation connects the gravitational potential, flow velocity, and thermodynamic properties, providing a comprehensive description of the system’s dynamics. In this framework, the inclusion of dark matter and stellar potentials significantly modifies the gravitational potential term $\Phi$, affecting both the structure and the position of the flow’s critical point.

\subsection{Gravitational Potential}\label{GP}

To provide a more physically motivated and accurate model of quasi-spherical accretion, we have applied a multi-component model of the Galactic potential. In cosmological simulations, the Bondi accretion rate is typically used to model the \ac{BH}'s gas accretion.

The total gravitational potential $\Phi$ is modeled as the sum of the potentials of the central black hole $\phi_{\rm BH}$, the dark matter halo $\phi_{\rm DM}$ and the stellar component of the galaxy $\phi_{\rm S}$ and can be expressed as

 \begin{equation}\label{phi_t}
\Phi = \phi_{\rm BH} + \phi_{\rm DM} + \phi_{\rm S}
\end{equation}

This formulation allows us to systematically investigate how different components of the Galactic potential influence accretion dynamics on large scales. To incorporate \ac{GR} effects within a Newtonian framework, we adopt the Paczy{\'n}sky–Wiita potential \citep{1980AA....88...23P}, which approximates the Schwarzschild geometry as:

\begin{equation}
\phi_{\rm BH} = -\frac{GM_{\rm BH}}{r - r_{\rm s}},
\end{equation}

where $r$ is the radial distance from the black hole, $G$ is the gravitational constant, $M_{\rm BH}$ is the mass of the black hole, and $r_{\rm s} = 2GM/c^{2}$ is the Schwarzschild radius (with $c$ being the speed of light). Pseudo-Newtonian potentials have limitations, as no single formulation accurately reproduces all dynamical properties of Schwarzschild space-time; a more precise alternative is the generalized Newtonian potential, which maintains a Newtonian framework while being derived directly from geodesic equations \cite{2013MNRAS.433.1930T}.

\begin{equation}
    \phi_{\rm G} = - \frac{G M_{\rm BH}}{r} - \left( \frac{r_{\rm s}}{1 - r_{\rm s}} \right) \left[ \left( \frac{r - \frac{r_{\rm s}}{2}}{r - r_{\rm s}} \right) \dot{r}^2 + \frac{1}{2} r^2 \dot{\phi}^2 \right]
\end{equation}

\begin{figure*}
    \centering
    \makebox[\linewidth]{%
\begin{tabular}{cc}
    \includegraphics[width=0.42 \linewidth]{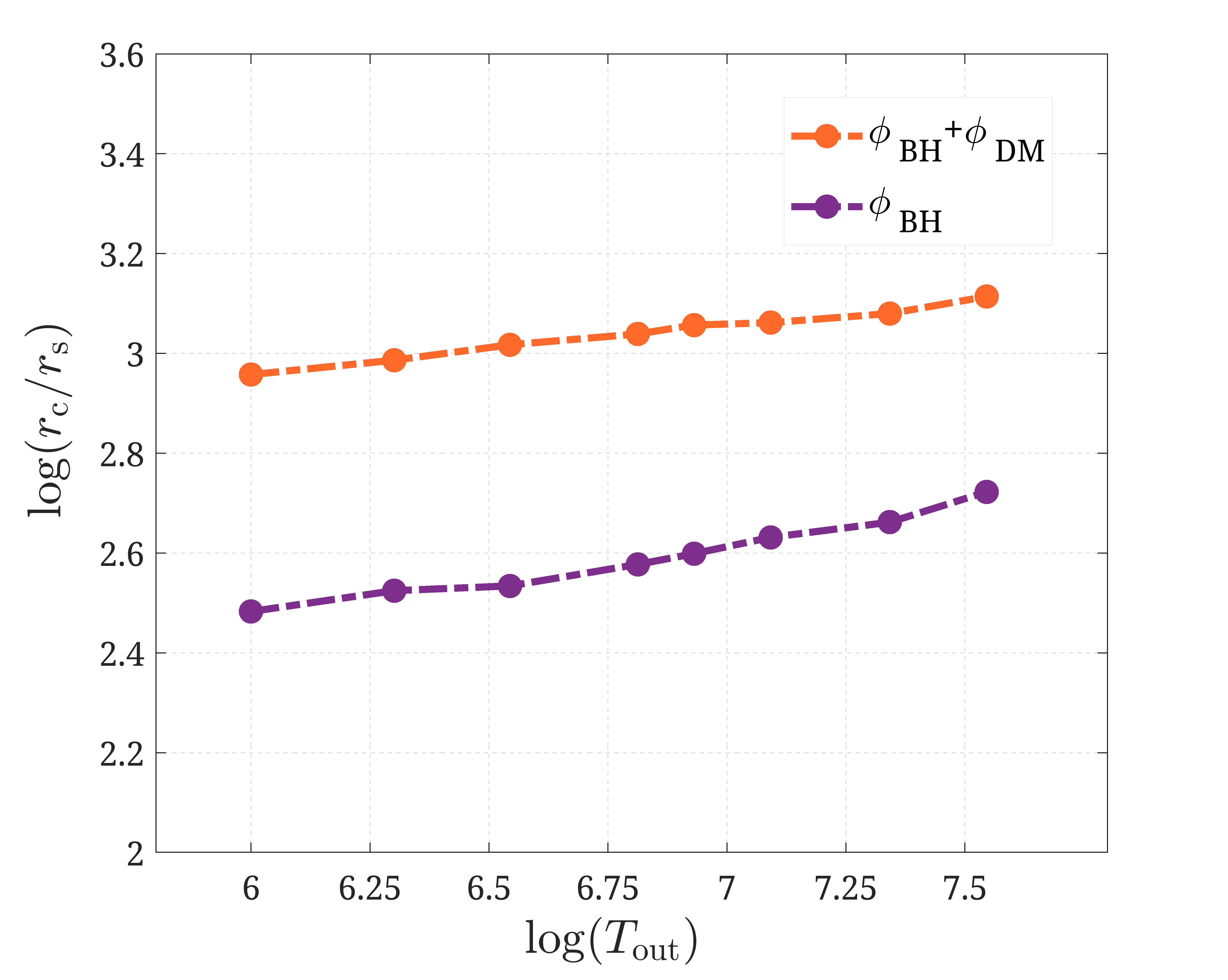}
     \includegraphics[width=0.42 \linewidth]{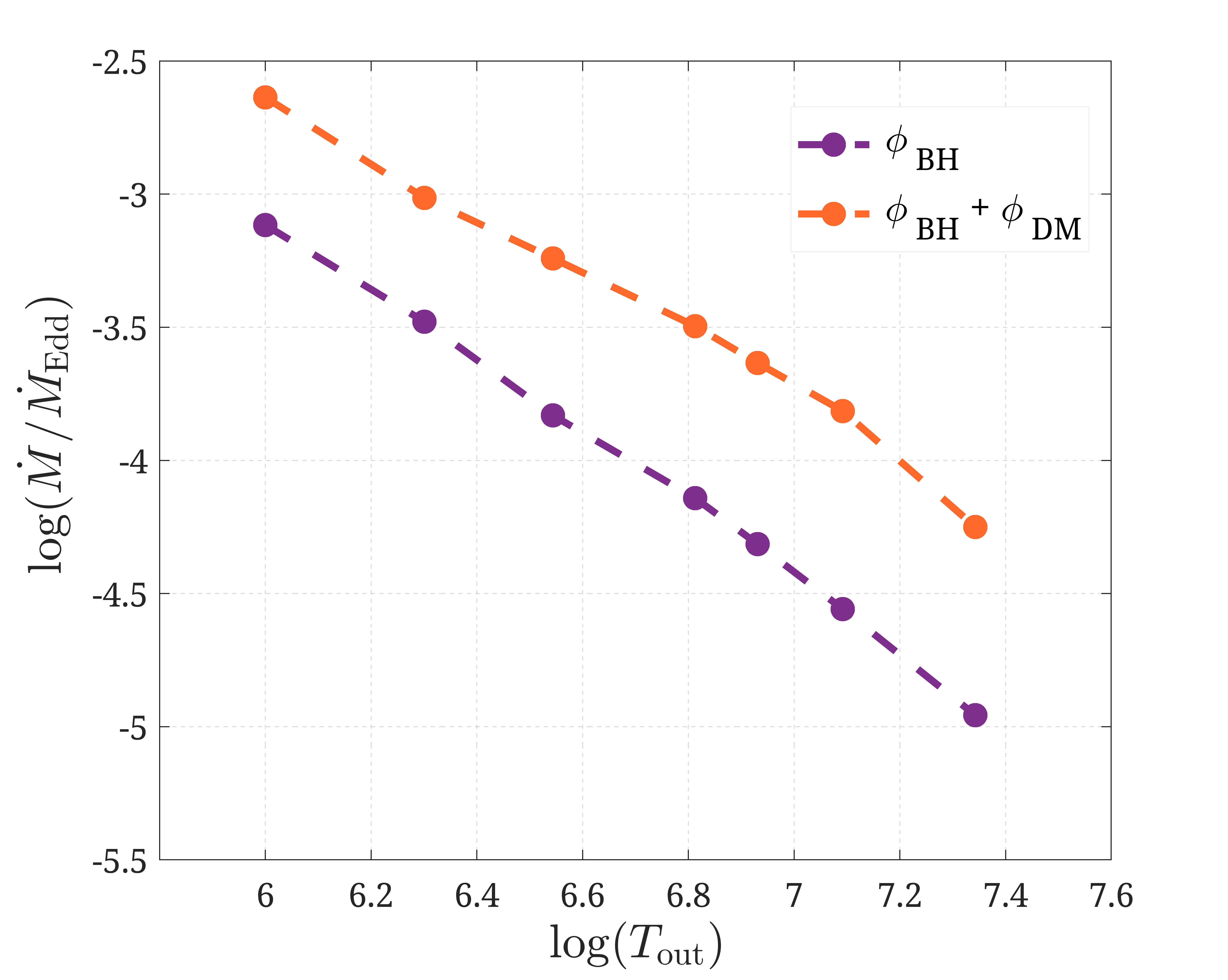}
           \end{tabular}%
  }
    \caption{Left panel: Position of the critical point as a function of the temperature at the outer boundary $T_{\rm out}$. Violet dots are cases without dark matter potential, and orange dots are cases with dark matter potential. Right panel: Mass accretion rate with zero angular momentum in units of Eddington accretion rate as a function of temperature at the outer boundary ($T_{\rm out}$). Violet dots show the mass accretion rate without dark matter potential, and orange dots show the mass accretion rate with dark matter potential. All solutions correspond to $a_{\rm 0} = 50,\ M_{\rm halo} = M_{\rm BH}$.}
    \label{fig3and4}

\end{figure*}

To assess the impact of these different potentials, we compared our results with those obtained using the generalized Newtonian potential (See Fig.\ref{multiphi} and section \ref{sec.4} for more details). 
The discrepancy in $\dot{M}$ between the two potentials was found to be relatively small. Given that our study primarily focuses on large-scale flow dynamics rather than the immediate vicinity of the black hole (i.e., \( r > 10\,r_{\rm s} \)) and as shown in Fig. \ref{P}, the effects of the generalized Newtonian potential are most significant only within $r<5\ r_{\rm s}$ ($\dot{r} = -0.141\ c, r\dot{\phi} = 0.038\ c $; see \cite{2013MNRAS.433.1930T} for more details). However, due to the complexities involved in this region and since our results focus on the effect of \ac{DM} halo on large-scale dynamics, we consider only regions beyond $10 \ r_\mathrm{S}$, where the PW and generalized Newtonian potentials are equivalent. Hereafter, $\phi_{\rm BH}$ refers to the PW potential.

Around the Bondi radius, the effect of the gravitational potential differs compared to the case of a pure black hole potential. Such a change can significantly alter the dynamics of accretion flow. Therefore, to study the accretion flow far away from the black hole, it is crucial to consider not only the gravitational potential of the black hole but also that of the galaxy stars and \ac{DM} halo, as they play significant roles and must be included. The second term of Eq. \ref{phi_t} pertains to the potential of the \ac{DM} halo, modeled using a Hernquist-type profile. The density distribution for this profile is given by:

\begin{equation}
\rho_{\mathrm{DM}} = \frac{M_{\rm halo}a_{\rm 0}}{2\pi r(r+a_{\rm 0})^3}
\end{equation}

where $M_{\rm halo}$ is the total mass of the halo, and $a_0$ is a typical length scale. The potential of the dark halo, assumed to be spherical, is:

\begin{equation}
 \phi_{\rm DM} = - \frac{GM_{\rm halo}}{r+a_{\rm 0}}
\end{equation}

where $G$ is the gravitational constant. This potential becomes significant at larger radii, where the \ac{DM} halo dominates over the stellar and \ac{BH} contributions, as shown in Figure \ref{P}.

We note that a \ac{BH} can alter the surrounding \ac{DM} distribution near the galactic center; however, this effect is not relevant for the purposes of the present study. Our focus here is on how the flow is influenced by the combined gravitational field of the galaxy (stars plus \ac{DM}) at large distances from the \ac{SMBH}. Moreover, because the halo’s response to the \ac{BH} potential can depend on the assumed \ac{DM} model, we adopt the conservative approach of keeping the \ac{DM} profile fixed. A full treatment of a dynamically evolving \ac{DM} profile lies beyond the scope of this work.

Also to assess the robustness of this approximation and incorporate more realistic physical conditions, we extend our analysis across a wide range of halo masses ($M_{\mathrm{halo}} = 1 - 10000 \ M_{\mathrm{BH}}$), including scenarios in which the halo mass greatly exceeds the central black hole mass—as is typical in galaxies.
For example, the Milky Way’s Sgr A* and the SMBH at the center of M87* are both embedded in halos with mass ratios on the order of $M_{\rm halo}/M_{\rm BH} = 10^4$, placing them near the upper bound of our examined range. 
 
\begin{figure*}
\centering
\makebox[\linewidth]{%
\begin{tabular}{cc}
    \includegraphics[width=0.42\linewidth]{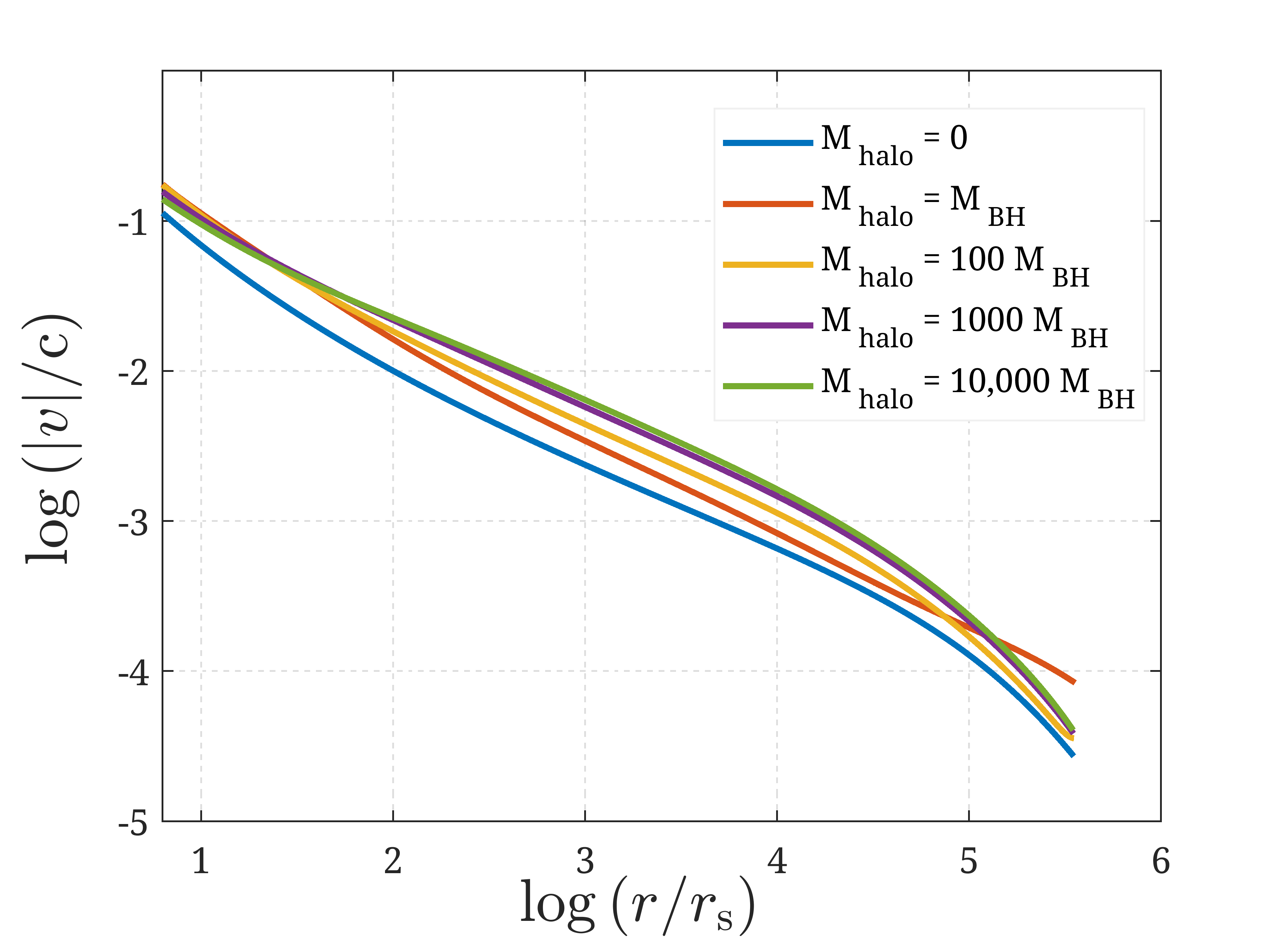}&
    \includegraphics[width=0.42\linewidth]{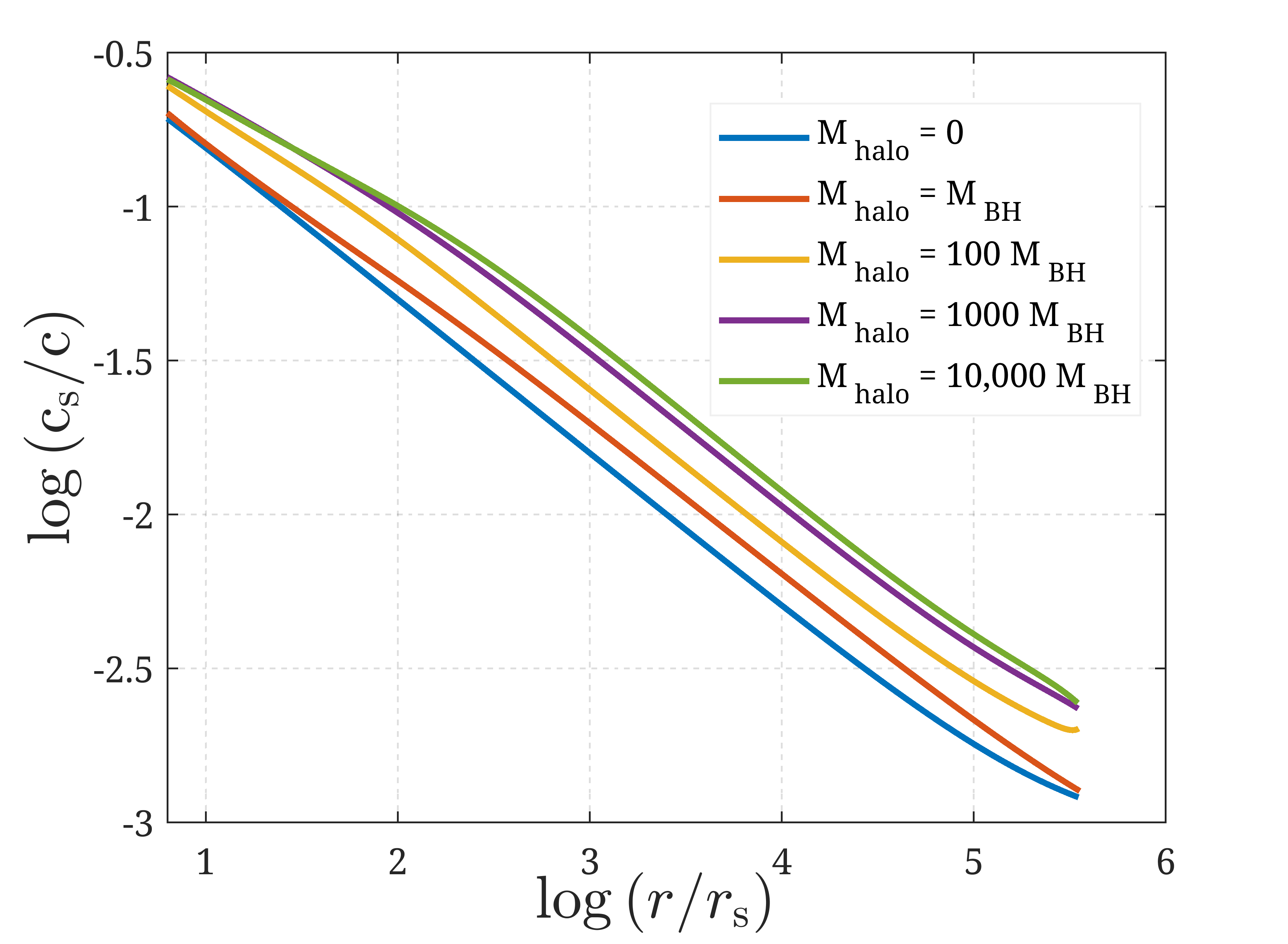}\\
    \includegraphics[width=0.42\linewidth]{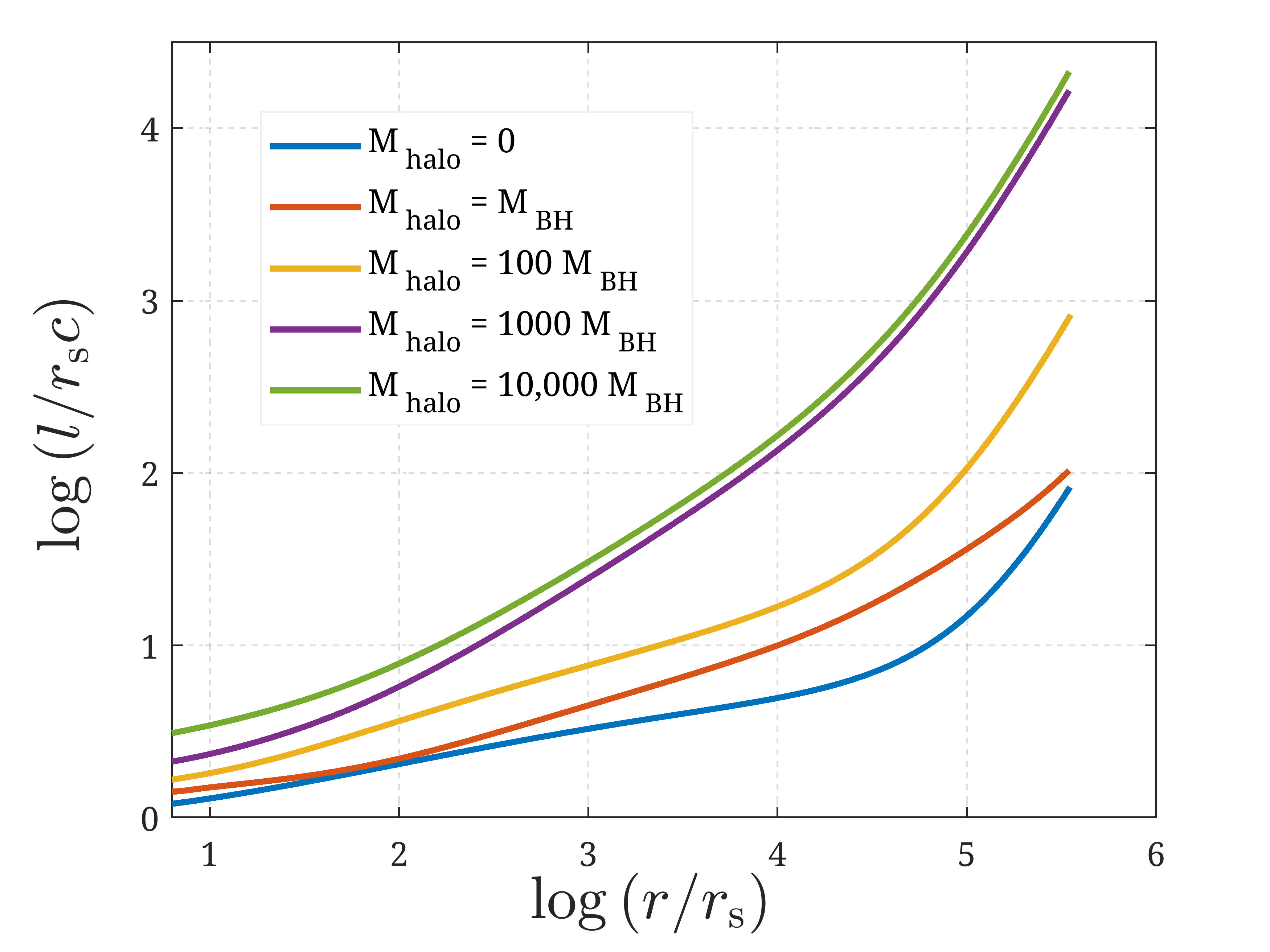}&
    \includegraphics[width=0.42\linewidth]{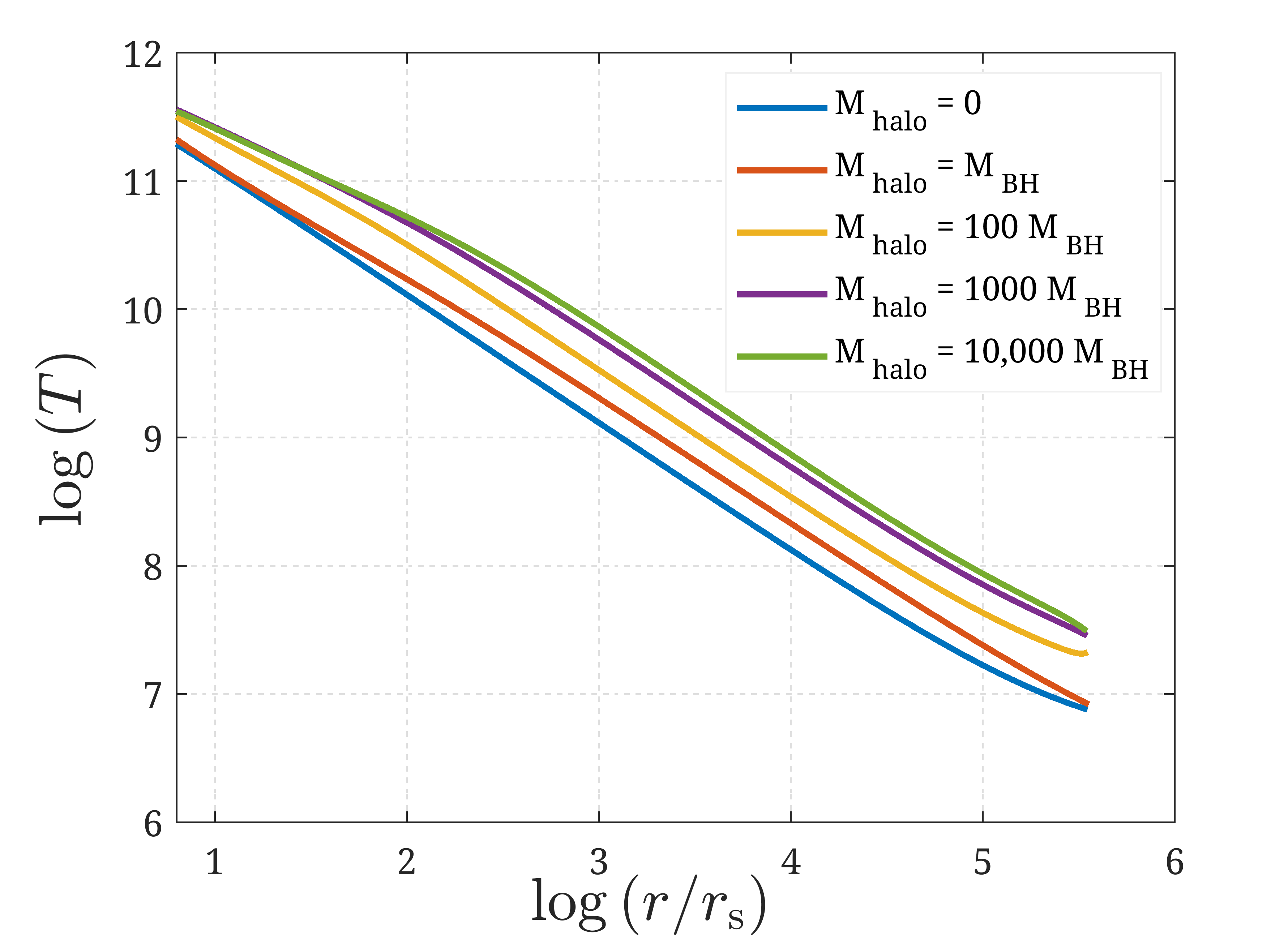}\\
  \end{tabular}%
  }
  \centering  
    \caption{Global solutions of slowly rotating accretion flows for different halo-to-black hole mass ratios. The blue line represents the case without dark matter. The red, yellow, violet, and green lines correspond to $a_{\rm 0} = 100$ and halo-to-black hole mass ratios of $M_{\text{\rm halo}}/M_{\text{\rm BH}} = 1$, 100, 1000, and 10000, respectively.}
    \label{2.1}
\end{figure*}

The third term of Eq. \ref{phi_t} represents the gravitational potential of the galaxy’s stars. Observations also show that the stellar and dark matter components of galaxies are distributed so that their total mass profile is well described by a density distribution $\rho \propto r^{-2}$ over a large radial range (e.g., \cite{2008MNRAS.384..987C}; \cite{2008MNRAS.388..384D}). Following \cite{2020ApJ...890...80C}, this motivates treating the galaxy potential as a flat rotation curve with a constant velocity dispersion parameter. Then, the gravitational potential of the stellar bulge is written as

\begin{equation}
\phi_{\rm S}(r) = \sigma^2 \ln r + C ,
\end{equation}

 Here, $\sigma$ is a constant parameter of the stellar potential and should not be directly equated with the observed one-dimensional stellar velocity dispersion. We assume $\sigma = 200$ km s$^{-1}$, which is commonly used for elliptical galaxies \cite{2013ARAA..51..511K}.

A key question is how the above gravitational potential influences the system at large values of $r$. Most numerical simulations of accretion flows have primarily focused on regions close to the black hole, making it difficult to investigate large-scale gravitational effects. To address this question, we need to study spherical accretion flows at large distances from the black hole, near or beyond the Bondi radius. At large distances, the gravitational potential of the galaxy's stars and the dark matter halo--in addition to the black hole's potential--becomes significant and must be taken into account. Around the Bondi radius, the slope of the gravitational potential differs from that of a pure black hole potential (see Figure \ref{P}). Such a change could have a significant impact on the dynamics of accretion.

Figure \ref{P} shows the magnitudes of the three potentials on the same graph to compare their values at different radii. Black hole potential dominates at very small scales, close to the Schwarzschild radius. The gravitational influence is very strong near the event horizon and quickly diminishes with distance. The green line represents the generalized Newtonian potentials. Its deviation from the Paczyński-Wiita potential is noticeable only at radii $r< 5 \ r_{\rm s}$. Dark matter halo potential becomes significant at larger scales, reflecting the impact of the dark matter distribution in the galaxy. It remains relatively flat over a wide range of distances compared to the black hole potential. This potential represents the gravitational influence of the distribution of stars within the galaxy. It shows a more gradual change with distance, usually logarithmic, reflecting the spread of stellar mass. Shaded regions in the figure highlight the domains where each potential dominates: the blue region indicates where the black hole potential is dominant, while the grey-shaded region indicates where the 
gravitational potential is dominated by the dark matter component. Near the Bondi radius, both the \ac{DM} and stellar potentials exceed the black hole potential.

\subsection{Numerical Method}
Spherical accretion has a critical point, which prevents the direct integration of the equations from the outer subsonic region to regions around $10 \ r_s$. This difficulty is typically addressed in the following way: either one starts with an arbitrarily chosen critical radius, integrates both inward and outward with the help of regularity conditions, and adjusts the critical radius until the outer and inner boundary conditions are satisfied. 
This approach ensures a smooth and physically consistent transition across the critical point. The governing differential equations are:

\begin{equation}
    \frac{dc_{\rm s}}{dr} = \frac{(\gamma - 1)}{2 v c_{\rm s}} \left( -\frac{2 v c_{\rm s}^2}{r} - c_{\rm s}^2 \frac{dv}{dr} \right)
\end{equation}

\begin{equation}
    \frac{dv}{dr} = \frac{v}{\gamma c_{\rm s}^2 - v^2} \left( \Omega_{\rm k}^2 r - \frac{2 \gamma c_{\rm s}^2}{r} + \frac{G M_{\rm halo}}{(r + a_{\rm 0})^2} + \frac{\sigma^2}{r} \right)
\end{equation}

where $\Omega_{\rm K}$ is the Keplerian angular velocity given by

\begin{equation}
   \Omega_{\rm K}^2 = \frac{GM}{r(r-r_{\rm s})^2}
\end{equation}
which corresponds to the orbital frequency derived from the Paczynski–Wiita potential.
The location of the sonic point is unknown and is an eigenvalue, so we need a set of three boundary conditions to solve the equations. Similar to the approach taken by \cite{2011MNRAS.415.3721N}, we solve the equations across multi-scale regions.

\subsection{Boundary Condition}

The dynamics of an astrophysical accretion flow can be characterized by a system of nonlinear ordinary differential equations. Consequently, the boundary condition assumes a potentially significant role in addressing this boundary value problem. Considering the inherent transonic nature of accretion onto a black hole, it is essential for the solution to fulfill the sonic-point condition at a distinct radius referred to as the sonic radius.

\begin{equation}
    \frac{dv}{dr} = \frac{N}{D}
\end{equation}

We impose two boundary conditions at the sonic radius and one at the outer boundary. The
boundary conditions at the sonic radius are as follows:

\begin{equation}
\gamma c_{\rm s}^2 - v^2 = 0, \quad r = r_c    
\end{equation}

\begin{equation}
\Omega_{\rm k}^2 r - \frac{2\gamma c_{\rm s}^2}{r} +\frac{GM_{\rm halo}}{(r+a_{\rm 0})^2} + \frac{\sigma^2}{r}= 0
\end{equation}

These conditions ensure that the flow transitions smoothly through the sonic point, where the flow velocity equals the local sound speed. This transonic behavior is essential for constructing physically realistic accretion models. Since we use a Bondi-like accretion model here, it is worth noting that, in the classical Bondi accretion onto a black hole in the adiabatic case with $\gamma=5/3$, there is no transonic behavior, as the sonic point coincides with the origin.

\begin{figure}
       \centering
    \includegraphics[width=1.05\linewidth]{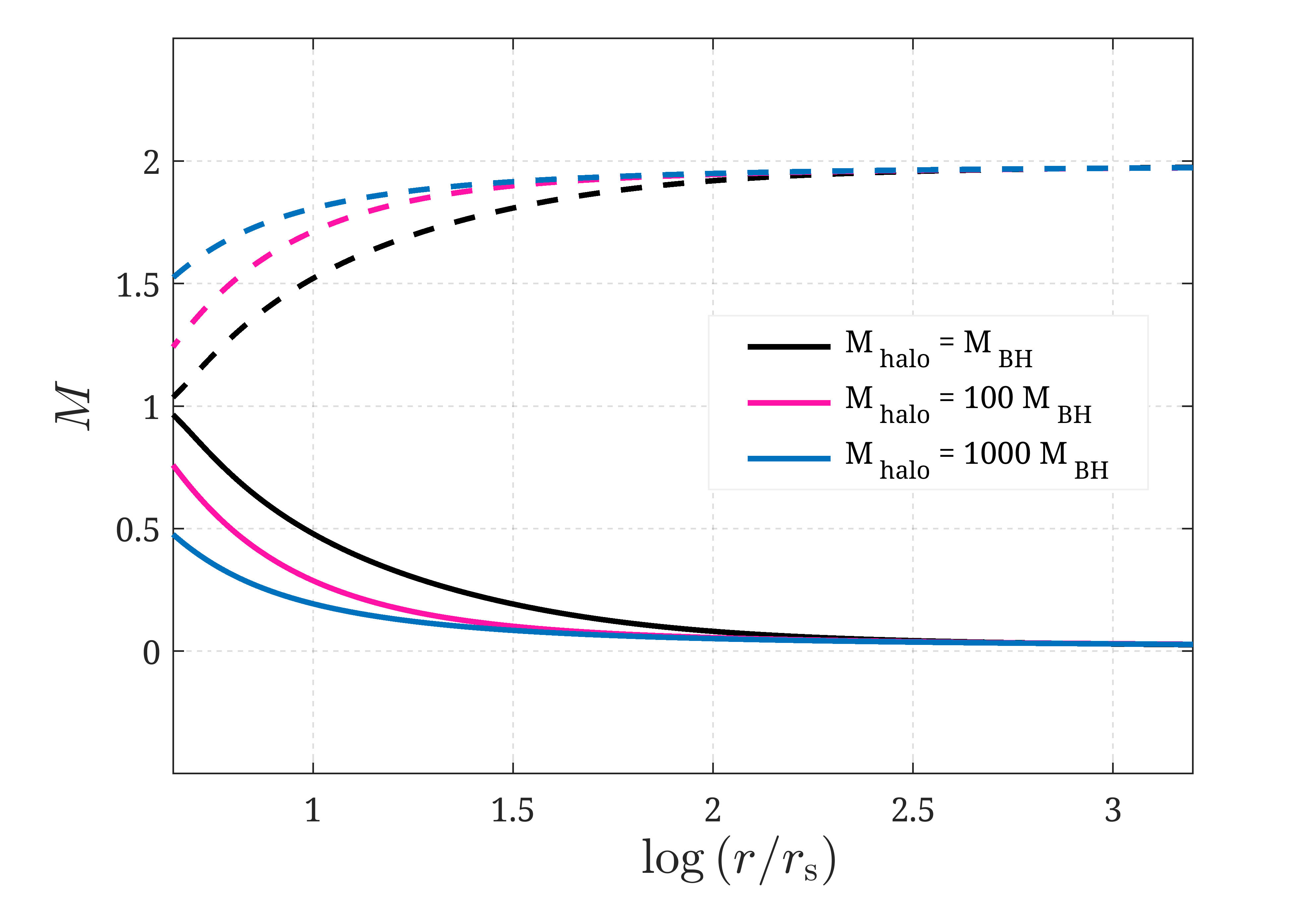}
    \caption{ Global numerical solutions for the Mach Number with ($ a_0 = 100,\ \mathcal{L} = 85$). The Black, pink and blue lines are for $ M_{\rm halo}/M_{\rm BH} = 1, 
 100$ and 1000 respectively.}
    \label{fig5}
\end{figure}

 On the other hand,  in complex astrophysical environments, the accreting gas at the outer boundary exhibits a variety of states, characterized by factors such as temperature and angular momentum. Consequently, an analysis of the accretion process necessitates consideration of the role played by the outer boundary condition (OBC). A realistic model must reflect the sensitivity of the accretion power to the conditions at the outer boundary. At the outer boundary

\begin{equation}
    c_{\rm s} = c_{\rm out}
\end{equation}

In our study, we designate the Bondi radius as the outer boundary in our solutions, which is $\sim 10^{5.7} \ r_{\rm s}$. While the classical Bondi model addresses accretion without angular momentum, realistic cosmic accretion flows inevitably involve some degree of rotation. Accounting for this rotation introduces additional complexity but provides a more realistic depiction of the process. We adopt geometrized units ($r_{\rm s} = c = 1$), and use $r_{\rm s}$ as our unit of length.

\section{Low angular momentum accretion}\label{sec.3}

The angular momentum of an accretion flow plays a crucial role in determining its flow characteristics. Low angular momentum accretion flow onto astrophysical black holes may manifest as practically inviscid flow in realistic astrophysical systems \cite{2023AA...678A.141O}, such as detached binaries fueled by accretion from OB stellar winds \cite{1975AA....39..185I}.
Additionally, recent studies focusing on accretion onto the black hole at the center of our galaxy have also provided evidence for the existence of such flows \citep{2006MNRAS.370..219M, 2006AA...450...93M}. These flows represent an intermediate case between
purely spherical accretion and disk accretion, making their study essential for understanding a wide range of
astrophysical systems.

The equations governing low-angular-momentum accretion flow are:

\begin{equation}
    \frac{d}{dr}(4 \pi \rho v r^2) = 0,
\end{equation}
\begin{equation}
    v\frac{dv}{dr} = r\Omega^2 - \frac{d\Phi}{dr} - \frac{1}{\rho}\frac{dp}{dr}
\end{equation}
\begin{equation}
    v \frac{d}{dr}(\Omega r^2) = \frac{1}{\rho r^2}( \rho \nu r^4 \frac{d\Omega}{dr})
\end{equation}
\begin{equation}
    \frac{\rho v}{(\gamma -1)} \frac{dc_{\rm s}^2}{dr} - c_{\rm s}^2 v \frac{d\rho}{dr} = \rho \nu r^2 (\frac{d\Omega}{dr})^2
\end{equation}

where $\nu$ is the kinematic coefficient of viscosity, which is defined by $ \nu = \alpha c_{\rm s} v$, the rotating accretion flow should have some viscosity that transports the angular momentum outward, thereby enabling the inward accretion of gas. $\Phi$ is the gravitational potential, as given in Eq.\ref{phi_t}, which includes Paczynski–Wiita potential for the black hole, the potential of stars, and the dark matter halo potential. To solve a differential equation in the subsonic region, we use a numerical technique similar to that described in \cite{2023ApJ...954..117R}. For additional details, see the appendix. 

\section{Numerical Results}\label{sec.4}

This paper aims to obtain numerically exact global solutions for accretion flows with zero angular momentum and low angular momentum in the presence of a \ac{DM} potential. Since self-similar solutions cannot describe flows in boundary regions, we utilize global solutions to understand the properties of the flow near the inner and outer boundaries. In this study, we employ the Hernquist profile to describe the spherical \ac{DM} halo in elliptical galaxies. In the following, we present the results of our numerical calculations and examine their physical implications. We pay particular attention to the effects of including the galaxy potential (DM + Stars) in the solutions. The parameters used in our solutions are $a_{\rm 0} = 50 $ and $M_{\rm halo} = M_{\rm BH}$.

\begin{figure}[b]
       \centering
    \includegraphics[width=1.1\linewidth]{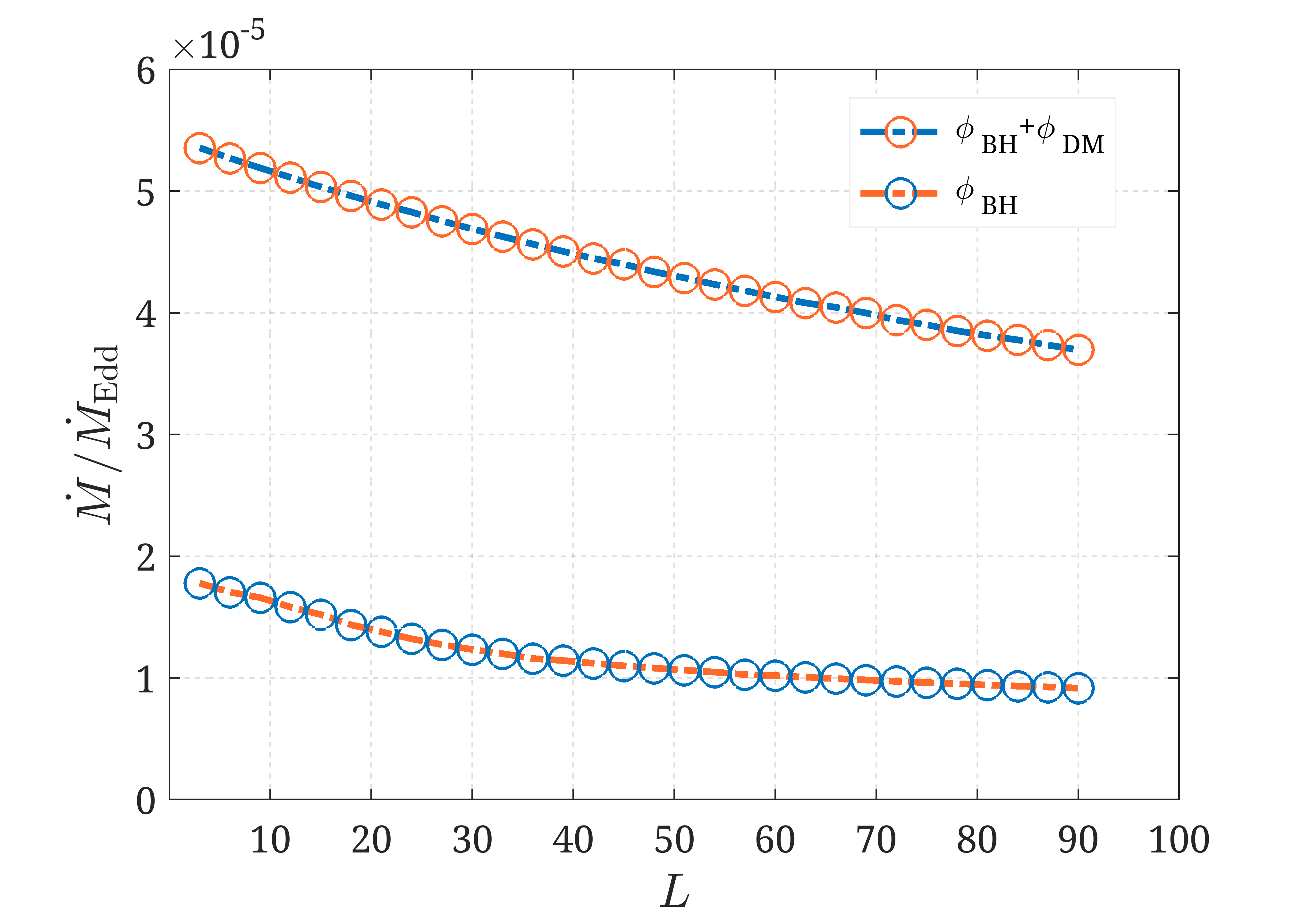}
    \caption{Mass accretion rate in units of Eddington accretion rate as a function of angular momentum at the outer boundary $\mathcal{L}>1$, The orange dots are for $a_{\rm 0} = 100,\ M_{\rm halo} = M_{\rm BH}$ and the blue line is the case without dark matter potential. }
    \label{fig6}
\end{figure}
\begin{table*}[ht]
\centering
\label{tab:models}
\begin{tabular*}{\textwidth}{@{\extracolsep{\fill}}lllllllll@{}}
\toprule
\hline
Model & Halo Gravity & Stars Gravity & $M_\text{\rm halo}$ & $a_{\rm 0}$ & $L$ & $T_\text{\rm out}$ & $\dot{M}_\text{\rm BH}/\dot{M}_\text{\rm Edd}$ \\
\midrule
Bondi & OFF & OFF & 0 & 0 & 0 & $6.5 \times 10^6$ & $7.21 \times 10^{-5}$ \\
Bondi & OFF & OFF & 0 & 0 & 0 & $1.0 \times 10^6$ & $7.65 \times 10^{-4}$ \\
Bondi & ON  & ON  & 1 & 50 & 0 & $6.5 \times 10^6$ & $5.12 \times 10^{-4}$ \\
Bondi & ON  & OFF & 1 & 50 & 0 & $6.5 \times 10^6$ & $3.19 \times 10^{-4}$ \\
Bondi & ON  & ON  & 1 & 100 & 0 & $6.5 \times 10^6$ & $3.00 \times 10^{-4}$ \\
\hline
SRAF  & OFF & OFF & 0 & 0   & 85 & $6.5 \times 10^6$ & $9.30 \times 10^{-6}$  \\
SRAF  & OFF & OFF & 0 & 0   & 85 & $1.0 \times 10^6$ & $2.01 \times 10^{-4}$  \\
SRAF  & ON  & OFF & 1 & 50  & 85 & $6.5 \times 10^6$ & $4.81 \times 10^{-5}$  \\
SRAF  & ON  & OFF & 1 & 100 & 85 & $6.5 \times 10^6$ & $3.75 \times 10^{-5}$  \\
SRAF  & ON  & OFF & 1 & 100 & 85 & $1.0 \times 10^6$ & $3.71 \times 10^{-4}$  \\
\bottomrule
\end{tabular*}
\caption{Summary of Models, Col 1, the type of accretion flow. Cols 2 and 3 correspond to DM halo gravity and star gravity, and col 4 is the mass of the dark matter halo in units of the black hole mass. Col 5 is a characteristic scale length related to the dark matter halo. Cols 6 and 7, Specific angular momentum and temperature at the outer boundary. Col 8, the mass accretion rate normalized to the Eddington accretion rate.}
\label{table}
\end{table*}

Figure \ref{1} displays sample solutions, including the radial profiles of velocity, sound speed, density, and pressure corresponding to $\alpha = 0.1$, $\gamma = 5/3$, $c_{\rm out} = 10^{-3}$ and $ \phi_{\rm galaxy} = 0 $. The solutions clearly fall within the Bondi regime, as the gas exhibits negligible specific angular momentum at the outer boundary. The radial velocity is significant but remains lower than the near-free-fall velocity characteristic of spherical Bondi flow. Notably, the angular momentum is zero in all solutions; they accurately represent a true Bondi accretion flow. The solution depicted by the upper curve in the four panels of Figure \ref{1} ( $\phi_{\rm DM} \neq 0 $) provides clear evidence of the influence exerted by the gravitational potential of dark matter. The sonic radius, $r_s$, indicated by the black dot, is located at $417 \ r_{\rm s}$ without a \ac{DM} halo and shifts to $745 \ r_{\rm s}$ when a dark matter halo is included. Figure \ref{1} demonstrates the effect of the \ac{DM} potential on Bondi accretion. In spherical accretion, the flow passes the critical point near the Bondi radius, becoming supersonic inside $r_{\rm c}$.

As shown in Figure \ref{fig3and4}, the left panel displays the position of the sonic point as a function of temperature at the outer boundary, revealing systematic shifts as the \ac{DM} potential is introduced. This result illustrates how the sonic radius varies under the influence of the \ac{DM} halo. This behavior is particularly interesting as it emphasizes the influence of \ac{DM} on the dynamics of accretion flows.
The right panel of Figure \ref{fig3and4} shows the variation of the mass accretion rate as a function of temperature at the outer boundary. These temperatures are representative of the interstellar medium (ISM) in a galactic nucleus. It is clear that the addition of the \ac{DM} potential significantly increases the mass accretion rate.

However, in the case of a rotating viscous accretion flow, a large portion of the flow remains subsonic well inside $r_B$ due to the rotational motion counterbalancing gravity. 

Figure~\ref{2.1} presents the radial variation of key dynamical
and thermodynamic quantities obtained from our solutions, showcasing how the structure of accretion flows
evolves under the influence of \ac{DM} halos of varying mass. The top panels depict the radial velocity and sound speed as functions of radius, while the bottom panels present the specific angular momentum, defined as $l = \Omega r^2$, and the temperature (Kelvin) of the flow. In this figure, the critical radius $r_c$ for a low angular momentum flow is much smaller compared to a zero angular momentum flow ($ r_{\rm c} < 10 \ r_{\rm s}$), which is out of the dominance of our solution. This highlights that our focus remains on larger, quasi-spherical inflow regions where the effects of external potentials are more pronounced.

We observe that when the halo mass is significantly larger than the black hole mass—such as in real systems like Sgr A* and M87*, where $M_{\mathrm{halo}} \sim 10000 \ M_{\mathrm{BH}}$—our solution improves slightly compared to the case where $M_{\mathrm{BH}} = M_{\mathrm{halo}}$. In this regime, the central black hole effectively acts as a small perturbation, and the Hernquist profile alone provides a good approximation to the full solution. Notably, the assumption of a toy-model \ac{DM} halo with $M_{\mathrm{BH}} = M_{\mathrm{halo}}$ remains valid and practical at small radii, especially when compared to more realistic cases with larger halo masses.

In addition, the presence of a \ac{DM} halo leads to a noticeable increase in both the specific angular momentum and the temperature of the accretion flow. This trend of increasing temperature with increasing halo mass is consistent with the fully general relativistic simulation results of \cite{2025AA...RH}, which shows that accretion flows become hotter in the presence of more massive dark matter halos. While their study 
models quasi-spherical accretion from $\sim 10$ pc down to the
event horizon using a fully general relativistic framework, our Bondi-like semi-analytic 
treatment is developed within a Newtonian context and focuses on the large-scale behavior of the flow.
These findings highlight the important role of the halo potential in shaping the flow structure,
particularly in low-angular-momentum environments.

Additionally, the low angular momentum flow also exhibits a lower radial infall velocity at the outer boundary due to the reduced centrifugal force (Figure \ref{1}). This correspondence of the low angular momentum flow to the spherical case reflect the fact that at the level of the basic equations of the hydrodynamics of the problem, angular velocity enters exclusively through quadratic terms, and hence, a perturbative angular momentum component will be ignorable to first order, appearing only to second order, as shown in \cite{2023ApJ...945...76H}.

Figure \ref{fig5} depicts the radial profile of the Mach number, showing changes in its structure in the presence of a \ac{DM} halo with different masses. These changes further underscore the impact of dark matter on accretion flow dynamics. Although the Mach number in \cite{2023ApJ...945...76H} remains constant with respect to radius, the overall trend shown in their Figure 2 is similar to the Mach number behavior observed in this work.

Finally, Figure \ref{fig6} presents the mass accretion rate in units of Eddington accretion rate as a function of angular momentum at the outer boundary, expressed in units of $L$. Our results indicate that, regardless of the value of $T_{\text{out}}$, the mass accretion rate $\dot{m}$ decreases as the rotation of the external gas $L$ increases, approaching the Bondi accretion rate as $L$ decreases. For the lowest angular momentum flows, the mass accretion rate is nearly equal to the Bondi rate, whereas for the highest angular momentum flows, it is approximately $20$ times lower than the corresponding Bondi rate. As seen in the figure, the effect of dark matter on Bondi-like accretion flow is more significant than in the case with low angular momentum.

Table \ref{table} compares different configurations of accretion models under various gravitational influences and parameters. The models include both Bondi and SRAF (Slowly rotating accretion flow) scenarios. Bondi Models: When both halo and star gravity are absent, the accretion rate is the lowest ( $7.21 \times 10^{-5}\ \dot{M}_{\rm Edd}$) and the critical radius is $ 386\ r_{\rm s}$.
Turning on star gravity (while keeping \ac{DM} halo gravity off) slightly increases the accretion rate. When \ac{DM} halo gravity is on (both with and without star gravity), the accretion rate increases significantly, and the critical radius also changes substantially. SRAF Models: Similar to the Bondi models, turning on halo gravity significantly increases the accretion rate.
The gravitational influence of stars also affects the accretion rate, though to a lesser extent than that of the \ac{DM} halo. 

Figure \ref{multiphi} illustrates the variation of the mass accretion rate as a function of $T_{\text{\rm out}}$. This figure compares the accretion rate of the classical Bondi model with a modified model that incorporates the gravitational potential of the galaxy, \ac{DM} halo, and rotation. At radii greater than $10\ \text{pc}$, the gravitational influence of the stars may surpass that of the black hole. The figure demonstrates that the inclusion of the galactic gravitational potential slightly increases the mass accretion rate. Furthermore, the introduction of slow rotation to the Bondi model results in a decrease in the mass accretion rate compared to the classical Bondi rate. The discrepancy between the generalized Newtonian potential (yellow dots) and the Paczyński-Wiita potential is minimal, showing a difference of only about $2\%$. In contrast, the effects of the stellar and \ac{DM} potentials differ from the Paczyński-Wiita potential by approximately $10 \%$ and $20\%$, respectively. Since our focus is on investigating the effects of the stellar potential and dark matter on large scales, we have chosen to disregard the region close to the black hole, where the generalized Newtonian potential can be important, and continue using the Paczyński–Wiita potential in our analysis.

\begin{figure}
       \centering
    \includegraphics[width=1.1\linewidth]{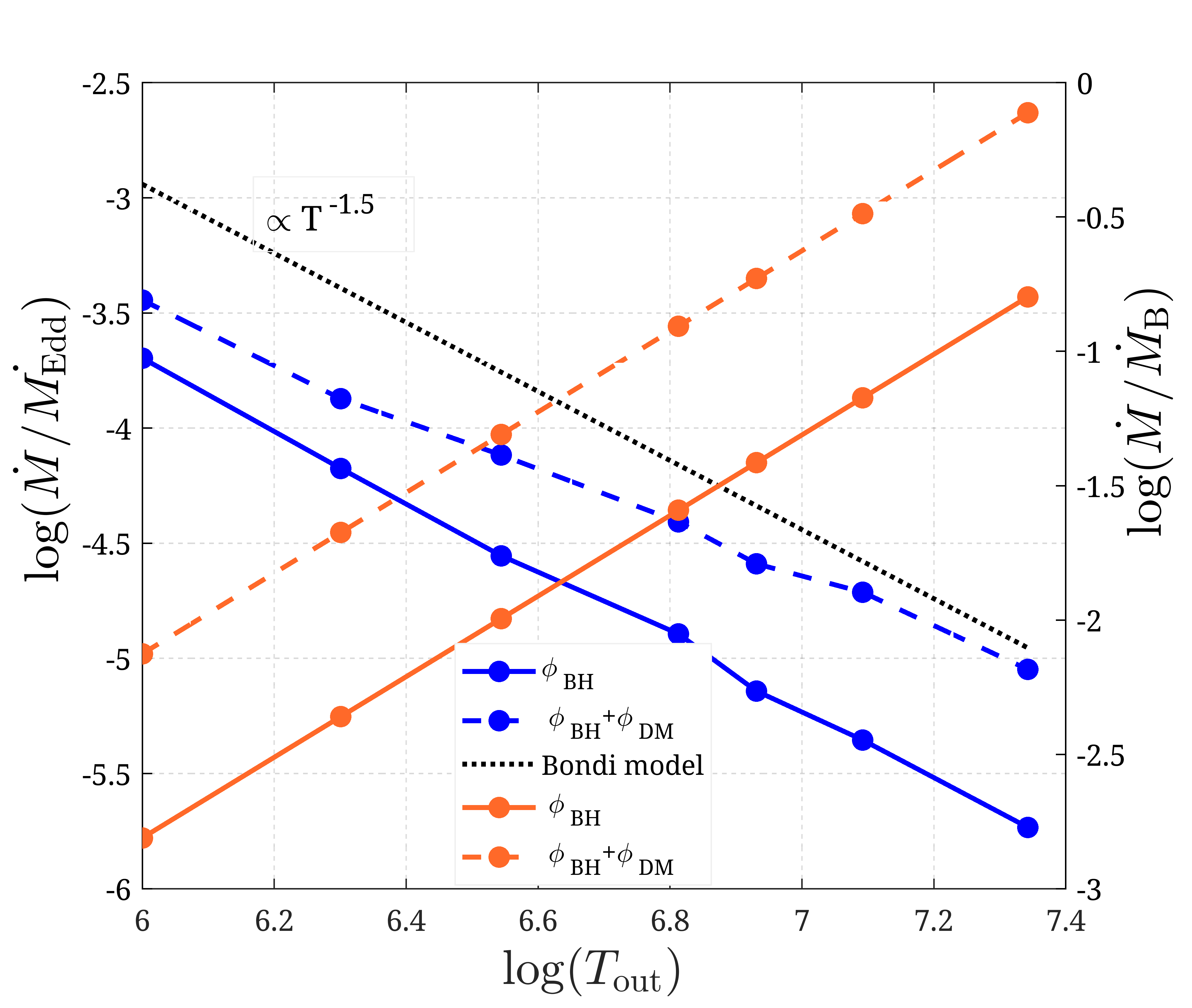}
    \caption{The dependence of accretion rates on the temperature at the outer boundary. We set the gas density at the outer boundary to be $10^{-24}\ gr\ cm^{-3}$. Red lines mean the accretion rates $\dot{M}/\dot{M}_{\rm Edd}$ predicted by our models, while blue lines mean $ \dot{M} / \dot{M}_{\rm Bondi} $. Black dotted lines mean the changing trend of the accretion rates predicted by the Bondi model.}
    \label{2yy}
\end{figure}

\begin{figure}
       \centering
    \includegraphics[width=1.1\linewidth]{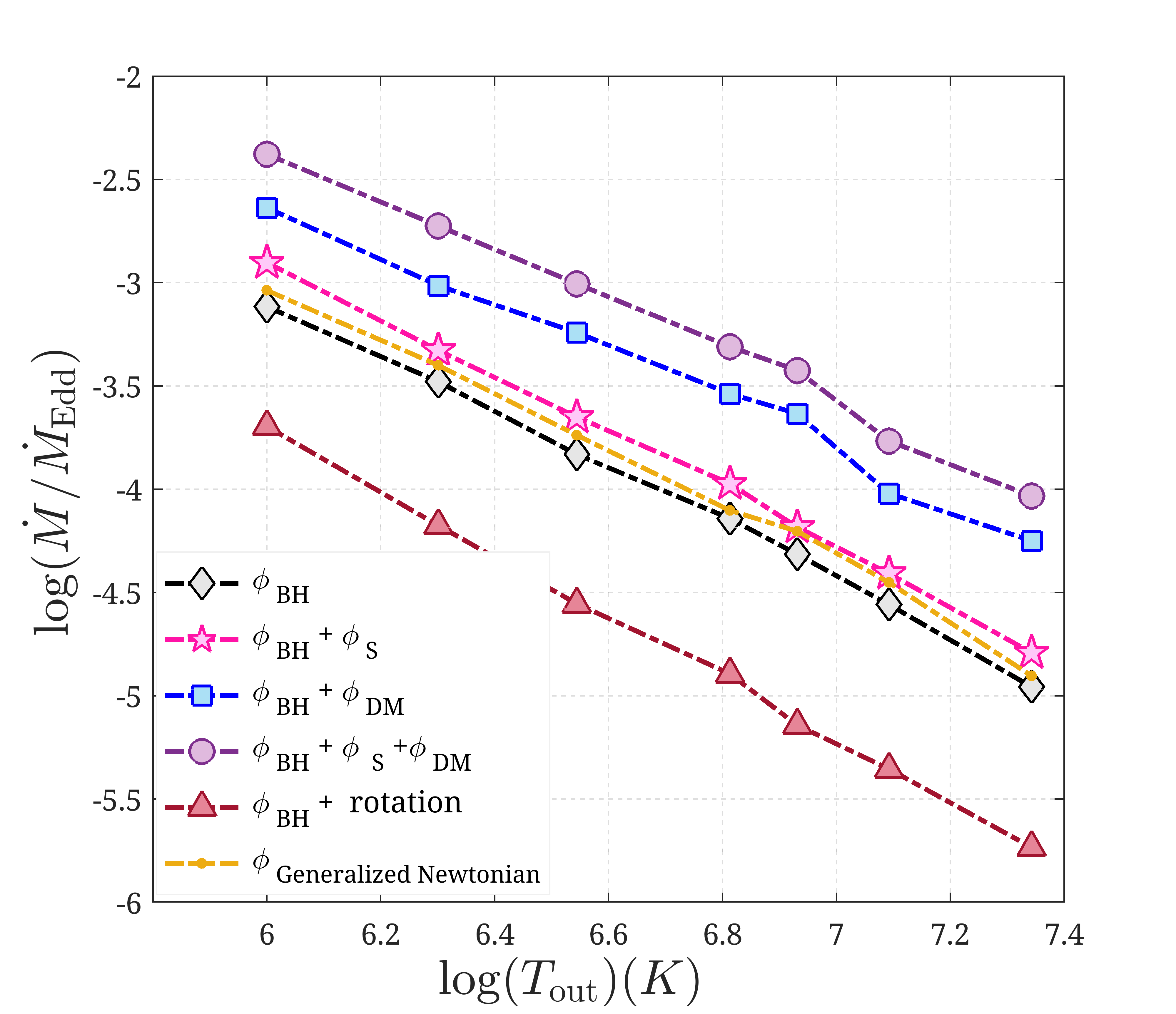}
    \caption{$M_{\rm halo} = M_{\rm BH}$, $a_{\rm 0} = 50$, Variations of accretion rate as a function of temperature at the outer boundary in different gravitational potentials and rotation. The black diamonds represent the accretion model with zero angular momentum in the Paczyński-Wiita potential of the black hole, while the yellow dots include the effects of the generalized Newtonian potential (\cite{2013MNRAS.433.1930T}). Pink stars are the Bondi model with galaxy potential. Blue squares include dark matter potential. Violet circles are both of them. Finally, red triangles show the accretion model with low angular momentum.}
    \label{multiphi}
\end{figure}

Our models have made modifications to the classical Bondi model, as depicted in Figure \ref{2yy}. This figure illustrates the dependence of accretion rates on temperature at the outer boundary $T_{\rm out}$. In Figure \ref{2yy}, the blue lines represent the accretion rates predicted by our model, normalized to the Eddington accretion rate, while the red lines indicate $\dot{M}/\dot{M}_{\rm Bondi}$. As shown in Figure \ref{2yy}, with the increase of $T_{\text{\rm out}}$, the accretion rates (blue lines) of our models decrease more slowly than the changing trend predicted by the Bondi model. Compared to our model, the Bondi model always overestimates the accretion rates, as shown by red lines. When the effect of the dark matter potential is included, the mass accretion rate is slightly increased. Overall, the gravitational potential of the galaxy contributes to an increase in accretion rates in our model.

These findings provide new insights into the behavior of accretion flows in the presence of dark matter, deepening our understanding of the complex dynamics in galactic nuclei and the role of dark matter in shaping these processes.

Several \ac{GR} simulations (e.g., \citep{2020ApJ...893...81T, 2025AA...RH}) have been conducted on spherical accretion flows, making it important to compare our results with these studies. One such work is \cite{2020ApJ...893...81T}, which presents a relativistic accretion model onto a Schwarzschild black hole. To validate our model, we compare some of our results—specifically those obtained without including the effects of dark matter and the galactic potential—with their findings.

Both studies examine how external factors influence accretion flow structure and mass accretion rates. We compare our dimensionless accretion rate $\dot{M}/\dot{M}_{\rm B}$ with \cite{2020ApJ...893...81T}, who use a full \ac{GR} framework and report $\dot{M}/\dot{M}_{\rm M}$ ($\dot{M}_{\rm M}$ is Michel accretion rate) specifically for an adiabatic index of $\gamma = 5/3$ and different sound speeds. Their results range from $0.98$ to $1.39$ for $c_{\rm s}=0.2$ to $c_{\rm s}=0.6$, while our values (without additional gravitational potentials) range from $0.01$ to $0.1$. 

Another simulation by \cite{2025AA...RH} examines the impact of a \ac{DM} halo on fully general relativistic spherical accretion flows. In the absence of dark matter, some of our solutions align with their findings, particularly their one-dimensional simulations at large distances from the black hole. Newtonian solutions have limitations, and including full GR effects are beyond the scope of this work. We plan to compare our Newtonian results with full GR simulations under similar conditions in future studies.

\section{DISCUSSION AND SUMMARY}\label{sec.5}

Given the extensive use of the classical Bondi theory, and considering that accretion onto \acp{SMBH} at the center of galaxies is certainly more complex than described by this theory—the motivation for this work was to generalize the theory, by incorporating the effects of a generalized gravitational potential due to a host galaxy (stars + dark matter). In section \ref{sec.2}, all the assumptions of classical Bondi accretion (stationarity, absence of rotation, and spherical symmetry) were maintained. In Section \ref{sec.3}, we consider an accretion flow with slow rotation in the presence of a galactic potential. The behavior of gas at the parsec scale can significantly impact the activity of low-luminosity active galactic nuclei (LLAGNs). In this study, we investigate the effect of dark matter on the properties of slightly rotating accretion flows around a non-rotating black hole. The model consists of a Schwarzschild black hole surrounded by dark matter, with a Bondi-like accretion flow. Initially, we consider spherically symmetric accretion flows at the outer boundary and introduce a small angular momentum to break the spherical symmetry. This study incorporates the Hernquist profile, a widely used model for \ac{DM} halos, into the Bondi accretion scenario to examine its influence on the structure of the accretion flow. This approach allows us to systematically analyze how dark matter affects the structure and dynamics of accretion flows. Although the role of \ac{DM} halos in Bondi accretion has been explored
in previous studies (e.g., \cite{2022MNRAS.512.2474M}), the present work is
distinct in that it incorporates the combined gravitational
influence of the central black hole, a stellar component,
and a dark matter halo modeled with a Hernquist profile.
This configuration is applied to a low-angular-momentum accretion flow,
allowing us to analyze the dynamics from galactic scales down to regions approaching
$\sim 10$ $r_{\rm s}$.
Throughout this study, we assume that the coupling between electrons and ions is sufficiently strong, resulting in a one-temperature plasma flow.

By introducing the \ac{DM} potential, we aim to explore its impact on the dynamics of the accretion flow, including the effects of small angular momentum. Our objective is to compare the influence of the \ac{DM} halo on both spherical accretion and slowly rotating accretion flows, gaining insights into the distinct characteristics and consequences of these scenarios in the presence of a \ac{DM} halo.

In agreement with previous
findings, we confirm that the presence of a \ac{DM} halo modifies the structure of Bondi-like accretion in
several key ways:

1. The inclusion of the dark matter halo, combined with the black hole potential, leads to a higher mass accretion rate. Dark matter enhances the gravitational pull at the galaxy's center, which in turn increases the mass accretion rate, consistent with previous studies. This enhancement may also play a crucial role in fueling LLAGNs, providing a pathway for interpreting observational data.

2. As previously discussed in \cite{2022MNRAS.512.2474M}, the \ac{DM} potential also affects the position of the sonic point. Our results confirm and extend their findings. In the Bondi accretion scenario, the presence of \ac{DM} causes a significant shift in the location of the sonic point. Also with an increase in temperature at the outer boundary, the sonic point, or sonic surface, may also approach to the Bondi radius.

3. In this work, as an extension of previous studies, we investigated low-angular-momentum accretion flows embedded within a Hernquist-profile \ac{DM} halo. Our results indicate that, for halos of comparable size and mass, the influence of \ac{DM} on low-angular-momentum accretion flows is less significant than on spherical accretion flows.

Our results indicate that the black hole mass, $M_{\mathrm{BH}}$, fundamentally sets the scale of accretion. Since the Bondi radius scales as $r_{\rm B} \propto M_{\mathrm{BH}}$, more massive black holes accrete from larger volumes where the influence of the host galaxy’s potential—including contributions from the stellar bulge and \ac{DM} halo—becomes increasingly significant. As a result, the displacement of the sonic point and the enhancement of accretion rates due to external potentials are more pronounced at higher $M_{\mathrm{BH}}$. 

While a full parametric study in $M_{\mathrm{BH}}$ is beyond the scope of this work, our framework lays the groundwork for future investigations into how black hole growth and galaxy-scale potentials jointly shape accretion flows across a broad mass range.

Future investigations could consider exploring additional \ac{DM} halo profiles, such as the NFW(Navarro-Frenk-White) and Jaffe profiles. Incorporating these alternative profiles into our analysis may provide valuable insights into the broader impact of dark matter on the accretion process, particularly in regions farther from the central black hole, where the \ac{DM} distribution may significantly influence the dynamics.

Additionally, it would be valuable to examine the influence of a \ac{DM} halo on outflow phenomena. Studying whether dark matter affects the intensity of outflows could provide critical insights into the complex relationship between dark matter and the overall dynamics of accretion processes.

\section*{Acknowledgements}
The authors would like to thank M. Roshan for his insightful comments and careful analysis regarding the limitations of the static Hernquist halo assumption. RR also sincerely thanks O. Porth for initiating the discussion that led to this work.
This work is supported by CIDMA under the Portuguese Foundation for Science and Technology (FCT, https://ror.org/00snfqn58)  Multi-Annual Financing Program for R\&D Units. HO is supported by the Individual CEEC program - 5th edition funded by the FCT, and acknowledges support from the projects PTDC/FIS-AST/3041/2020, CERN/FIS-PAR/0024/2021, and 2022.04560.PTDC. This work has further been supported by the European Horizon Europe Staff Exchange (SE) programme HORIZON-MSCA-2021-SE-01 Grant No. NewFunFiCO-10108625.

\appendix
\twocolumngrid

\section{Technical Method}\label{app}

The following section addresses the two-boundary value problem with unknown parameters, [$ j, r_{ \rm c} $].
Since the black hole accretion is necessarily transonic due to the nature of gravity around central objects, we define the critical point of the accretion flow in the following way: In principle, the critical point is a point of discontinuity of the differential equation and mathematically is defined as the $ \mathrm{d} v / \mathrm{d} r \to  0/0 $ form. 
Here, $ r_{\rm c} $ denotes the location of the critical point. In the following subsection, aim to simplify the system of equations to be solved in a range $ x = [ r_{\rm c}, x_{\rm out}] $.

\subsection{System of BVP equations}

The system of equations adopted in Section \ref{sec.3} is written as: 

\begin{equation} \label{dv}
	 \diff{v}{r} = f_{1r} =  v \frac{\mathcal{N}}{\mathcal{D}},
\end{equation}

\begin{equation} \label{domg}
	\diff{\Omega}{r} = f_{2r} = \frac{ v \left( \Omega r^{2} - j \right) }{\alpha r^{3} c_\mathrm{s} },
\end{equation}

\begin{equation} \label{cs}
	 \diff{c_{\rm s}}{r} = f_{3r} = \frac{\gamma-1}{2 v c_{\rm s}} \left[\frac{ v^{2} \left( \Omega r^{2} - j \right)^{2}}{\alpha  r^{3} c_{\rm s} } 
	- \frac{2 v c_{\rm s}^{2}}{r} - c_{\rm s}^{2} \diff{v}{r} \right],
\end{equation}

or equivalently,

\begin{equation} \label{cs2}
	f_{3r} = \frac{\gamma-1}{2 v c_{\rm s}} \left[\frac{ v^{2} \left( \Omega r^{2} - j \right)^{2}}{\alpha  r^{3} c_{\rm s} } 
	- \frac{2 v c_{\rm s}^{2}}{r} - c_{\rm s}^{2} f_{1r} \right],
\end{equation}

where,

\begin{align}
\mathcal{D} &= \gamma c_s^2-v^2,\\
\mathcal{N} &= \left( \Omega_\mathrm{_K}^{2} - \Omega^{2} \right) r - \frac{ 2 \gamma c_{\rm s}^{2}}{r} \notag \\
&\quad + \frac{(\gamma-1) \left( \Omega r^{2} - j \right)^{2} v}{\alpha r^3 c_{\rm s}} + \frac{GM_{halo}}{(r+a)^2} + \frac{\sigma^2}{r}.
\end{align}
    
The Keplerian angular velocity is given by:

\begin{equation}
	\Omega_\mathrm{_K}^{2} = \frac{GM}{r \left( r - r_\mathrm{s} \right)}.
\end{equation}

To determine the behavior of transonic solutions near the critical point, one needs to study $\left( \frac{\mathrm{d}v}{\mathrm{d}r} \right)_{\rm c}$. 
Since $\frac{\mathrm{d}v}{\mathrm{d}r} = \frac{N}{D} = \frac{0}{0}$ at a critical point, one must apply l’Hopital’s rule.

Since there is a singularity event near the horizon, we cannot use the direct integration of hydrodynamics equations (Runge-Kutta method) from the outer boundary toward the horizon to obtain smooth global solutions. Therefore, we employ a powerful technique called the iterative relaxation method \cite{1992...} to solve this problem. Although a simpler shooting method could be used, since our computational domain for realistic external media is $r_{out} \sim $~ parsec, it is preferable to use the relaxation method. Our problem is two-point boundary values, which means we do not know the values of variables at boundaries, so we need to use the guess function. This method begins with a guess, and as the solution is improved iteratively, the result tends to converge to the real solution. In this method, $j$ and $r_{\rm c}$ are eigenvalues so consider an initial guess for them and then integrate from the outer boundary to the sonic radius. For the inner region, we have algebraic equations, which can be easily solved. In our system, only one boundary abscissa ($r_{\rm out} = x_{\rm max}$) is specified, while the other boundary $r_{\rm c}$ is to be determined.

\subsection{Adding an extra constant dependent variable}

In the above system, only one boundary abscissa $ x_{\rm max} $ is specified, while 
the other boundary $ r_{\rm c} $ is to be determined. Therefore, we add an extra constant-dependent variable, as described by \cite{1992...},

\begin{equation}
	y_{4} \equiv \ln(r_{\rm c}) - \ln(x_{\rm max}),
\end{equation}

with the derivative:

\begin{equation}
	\diff{y_{4}}{r} = 0.
\end{equation}

We also define a new independent variable, $t$, by setting

\begin{equation}
	\ln(x) - \ln(x_{\rm max}) = t y_{4}, \quad \quad 0 \le t \le 1.
\end{equation}

Note that we choose $\ln(x)$ since the outer boundary is very far from the inner boundary, 
i.e., $ x_{\rm max} = r_\mathrm{_B} = 10^{5.7} r_{\rm s} $.

\subsection{Writing {\tt sys\_bvp.m} subroutine}
	The above system with a new independent variable $t$, can be written as,

\begin{equation}
	\diff{v}{t} =  \diff{v}{\ln x} \diff{\ln x}{t} = f_{1} = x y_{4} * f_{1r},
\end{equation}	
	
\begin{equation}
	\diff{\Omega}{t} = f_{2} = x y_{4} * f_{2r},
\end{equation}

\begin{equation} \label{dcs}
	 \diff{c_{\rm s}}{t} = f_{3} = x y_{4} * f_{3r},
\end{equation}

\begin{equation} \label{y}
	 \diff{y_{4}}{t} = f_{4} = 0,
\end{equation}

\subsection{Boundary conditions}

Point $ a $ corresponds to $ r_{\rm out} $, and point $ b $ shows the location of $ r_{\rm c} $ which 
correspond to $t = 0$ and $t = 1$, respectively.

At $ t = 0 $,

\begin{equation}
v_{a} = ya(1), \qquad \Omega_{a} = ya(2), \qquad c_{\rm sa} = ya(3),\\
\end{equation}
\begin{equation}
g(1)  = \Omega_{a} - \frac{L}{0.136} (\frac{c}{c_\mathrm{out}})^{-4},\\
\end{equation}
\begin{equation}
g(2) = c_{\rm sa} - c_\mathrm{out}.\\
\end{equation}

At $ t = 1 $,

 \begin{equation}
 v_{b} = yb(1), \qquad \Omega_{b} = yb(2),
\end{equation}	
\begin{equation*}
 c_{\rm sb} = yb(3), \qquad y_{4} = yb(4), 
\end{equation*}	

\begin{equation}
 r_{\rm c} = \exp( \ln(x_{\rm max}) + y4 )
\end{equation}	

\begin{align}
g(3) &= \left( \Omega_{\rm Kb}^2 - \Omega_{b}^2 \right) r_{\rm c} - \frac{2 \gamma c_{\rm sb}^2}{r_{\rm c}}, \\
&\quad + \frac{( \gamma - 1 )( \Omega_{b} r_{c}^2 - j )^{2}v_{b}}{\alpha r_{c}^3 c_{\rm sb}} + \frac{GM_h}{(r_{\rm c} + a)^2} + \frac{\sigma^2}{r_{\rm c}}\nonumber, \\
\end{align}	

\begin{equation}
g(4) = \gamma c_{\rm sb}^{2} - v_{b}^{2},
\end{equation}	
\begin{equation}
g(5) = \Omega_{b} r_{c}^{2} - j + \frac{2 \alpha c_{\rm sb} \Omega_b r_c^2}{v_b}.
\end{equation}

\bibliographystyle{aasjournal}
\bibliography{sample631}

\end{document}